\documentclass{article}

\usepackage{prime_arxiv}

\usepackage[utf8]{inputenc}
\usepackage[T1]{fontenc}
\usepackage{amsmath,amssymb}
\usepackage{graphicx}
\usepackage{subcaption}
\usepackage{cite}
\usepackage{hyperref}
\usepackage{siunitx}
\usepackage{booktabs}
\usepackage{tabularx}
\usepackage{multirow}
\usepackage{array}

\title{Predicting Grain Growth Evolution Under Complex Thermal Profiles with Deep Learning through Thermal Descriptor Modulation}

\author{Pungponhavoan ~Tep\thanks{Corresponding author: pungponhavoan.tep@minesparis.psl.eu} ,
Marc ~Bernacki \\
	\\
	Mines Paris, PSL University\\
 	Centre for material forming (CEMEF), UMR CNRS\\
 	 06904 Sophia Antipolis, France\\
}

\date{}

\begin{document}

\maketitle


\begin{abstract}
Predicting microstructure evolution during thermomechanical treatment is essential for determining the final mechanical properties of a material, yet conventional simulations based on Partial Differential Equations (PDEs) remain computationally expensive. Our prior Deep Learning (DL) framework using Convolutional Long Short-Term Memory (ConvLSTM) has proven effective in accelerating grain growth prediction, though its applicability was limited to constant-temperature or single-rate thermal profiles. As the model was trained exclusively under constant thermal conditions, it cannot account for the thermal history dependence of grain boundary kinetics, fundamentally limiting its applicability to the time-varying thermal profiles characteristic of industrial heat treatment processes. This study extends the previous framework by incorporating Feature-wise Linear Modulation (FiLM) for thermal conditioning to predict grain growth under complex, time-varying thermal profiles. The model was trained on a large dataset of grain growth evolution under thermal profiles with heating and cooling rates ranging from \SI{0.01}{\kelvin\per\second} to \SI{10}{\kelvin\per\second}. The results demonstrate that the proposed thermal conditioning mechanism enables the model to capture the influence of variable thermal profiles on grain boundary migration kinetics. Across the three test scenarios of increasing complexity, the model achieved a Structural Similarity Index Measure (SSIM) of up to \num{0.93} and mean grain size error ($\overline{R}$ error) below \SI{3.2}{\percent}. Despite the architectural extensions, inference time remains on the order of seconds per prediction sequence, preserving the computational advantage over PDE-based simulations.
\end{abstract}


\section{Introduction}

The evolution of microstructure during thermal processing determines the final mechanical properties of polycrystalline materials. Among relevant mechanisms, grain growth is driven by the reduction of grain boundary energy and governs the microstructural outcome of heat treatment processes~\cite{rohrer2023grain,hansen2004hall}. Conventional Partial Differential Equation (PDE)-based simulations, including front-tracking~\cite{florez2020novel,florez2021parallelization,florez20212d,mora2008three,barrales2010vertex}, phase-field~\cite{yang2021phase,moelans2008quantitative,krill2002computer,chen2002phase,steinbach2009phase}, and level-set methods~\cite{bernacki2024kinetic,li2025accurate}, provide accurate descriptions of microstructure evolution but incur substantial computational cost due to their sequential timestep calculations. Our previous study~\cite{TEP2025121486} addressed this limitation through a Deep Learning (DL) framework based on an encoder-decoder architecture coupled with Convolutional Long Short-Term Memory (ConvLSTM), achieving up to $90\times$ speedup over our in-house front-tracking simulation ToRealMotion (TRM)~\cite{florez2020novel,florez2021parallelization} while maintaining high-fidelity predictions, with a mean grain size error ($\overline{R}$ error) of \SI{0.07}{\percent} and a Structural Similarity Index Measure (SSIM) of \SI{86.71}{\percent} under isothermal grain growth conditions.

However, that framework was trained and evaluated exclusively under constant thermal conditions. In practice, industrial heat treatment involves complex sequences of heating, holding, and cooling stages with varying thermal rates. Since grain boundary migration kinetics depend on temperature through Arrhenius-type mobility~\cite{CHEN2020412}, an identical microstructure can evolve along different trajectories depending on the imposed thermal profile, which the previous model could not capture without explicit thermal descriptors.

Accordingly, this study extends the previous DL framework to non-isothermal grain growth prediction through Feature-wise Linear Modulation (FiLM)~\cite{perez2018film} thermal conditioning. By providing instantaneous temperature ($T$) and thermal rate ($\mathrm{d}T/\mathrm{d}t$) as explicit conditioning inputs alongside microstructure images, the model adapts its predictions to varying thermal conditions while preserving the computational efficiency of the original architecture. Performance was evaluated across three test scenarios of increasing complexity, from a simple heating-holding-cooling cycle to a complex multi-cycle thermal profile excluded from training, to assess both prediction accuracy and generalization capability.


\section{Methodology}

\subsection{Model Architecture}

\begin{figure}[h!]
    \centering
    \begin{subfigure}[t]{\linewidth}
    \centering
    \includegraphics[width=0.95\linewidth]{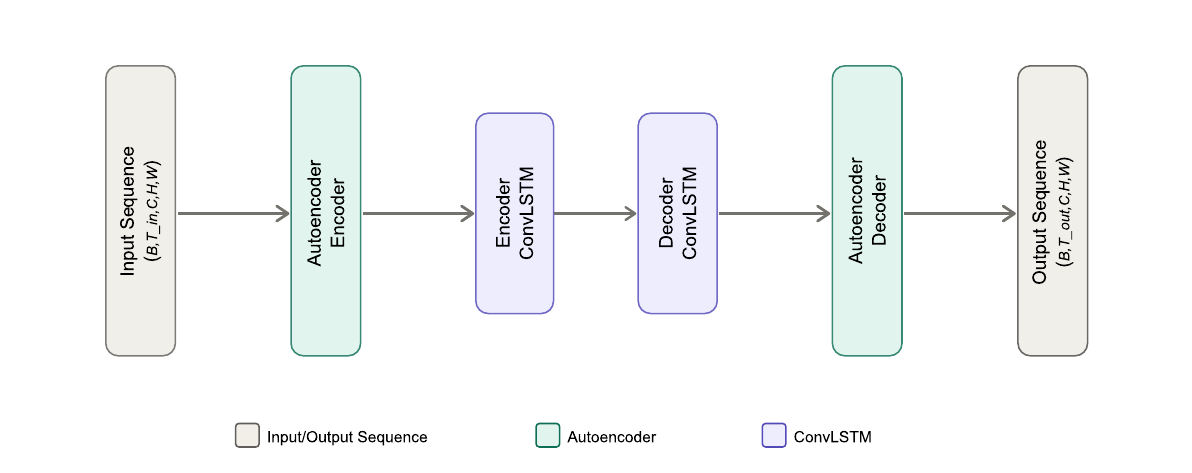}
    \caption{}
    \label{fig:arch_old}
    \end{subfigure}

    \vspace{0.75em}

    \begin{subfigure}[t]{\linewidth}
    \centering
    \includegraphics[width=0.95\linewidth]{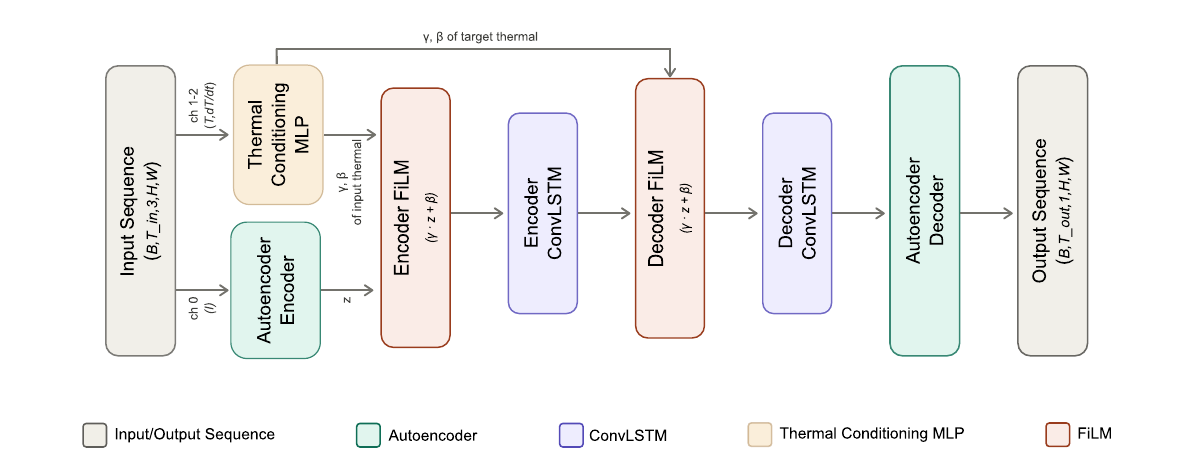}
    \caption{}
    \label{fig:arch_new}
    \end{subfigure}
    \caption{Neural network architectures for grain growth prediction: (a) encoder-decoder + ConvLSTM framework from the previous study; (b) extended framework with thermal conditioning through FiLM, where a thermal conditioning module generates scale ($\gamma$) and shift ($\beta$) parameters to modulate the latent features based on instantaneous $T$ and $\mathrm{d}T/\mathrm{d}t$.}
    \label{fig:arch}
\end{figure}

The framework proposed in the previous study, as illustrated in Figure~\ref{fig:arch}(a), consists of a Convolutional Autoencoder coupled with ConvLSTM, which compress microstructure images into a latent space, model spatiotemporal grain growth patterns, and reconstruct full-resolution predictions autoregressively. In that framework, only sequences of microstructure images were accepted as input, with boundary migration kinetics implicitly learned from constant-rate training data.

To extend prediction capability to complex thermal profiles, the architecture is augmented with a thermal conditioning mechanism through FiLM~\cite{perez2018film}, as illustrated in Figure~\ref{fig:arch}(b). At each timestep, $T$ and $\mathrm{d}T/\mathrm{d}t$ are processed through a Multi-Layer Perceptron (MLP) to generate scale ($\gamma$) and shift ($\beta$) parameters that modulate the latent features according to:
\begin{equation}
    \mathbf{z}' = \gamma \odot \mathbf{z} + \beta,
\end{equation}
where $\mathbf{z}$ and $\mathbf{z}'$ are the original and modulated latent representations, respectively. This element-wise affine modulation allows the model to adapt grain boundary migration kinetics dynamically in response to instantaneous thermal conditions throughout both input encoding and autoregressive prediction, without modifying the core encoder-decoder architecture.

\subsection{Dataset Generation}

The identical 18 initial microstructure states used in the previous study were retained, spanning a range of mean grain sizes, domain sizes, and topological characteristics representative of polycrystalline microstructures, as listed in Table~\ref{tab:dataset_gen_parameters_lavogen}. For each initial state, grain growth evolutions were simulated using the in-house TRM~\cite{florez2020novel,florez2021parallelization} under different thermal profiles of heating, isothermal holding, and cooling phases at thermal rates between \SI{0.01}{\kelvin\per\second} and \SI{10}{\kelvin\per\second}, as shown in Figure~\ref{fig:thermal_profiles} and listed in Table~\ref{tab:dataset_gen_parameters_trm}. The grain boundary mobility, $\mu$, is defined through an Arrhenius law:
\begin{equation}\label{Arrhenius}
\mu=\mu_0\exp\left(-\frac{Q}{RT}\right),
\end{equation}
with $\mu_0$ a constant pre-exponential term, $Q$ the activation energy, $R$ the gaz constant and $T$ the absolute temperature. The grain boundary energy $\gamma$ is assumed constant. The physical values used for $\mu_0,\ Q,\ \text{and}\ \gamma$ are also summarized in Table~\ref{tab:dataset_gen_parameters_trm} and are representative of the 304L stainless steel~\cite{Maire2017}.
Each sequence spanned \SI{59}{\minute} with microstructure states captured at \SI{1}{\minute} intervals alongside the corresponding $T$ and $\mathrm{d}T/\mathrm{d}t$ values computed as backward finite differences. After applying the same data augmentation techniques as the previous study, the final dataset contained \num{6727} evolution sequences, partitioned into training, validation, and test sets at \SI{80}{\percent}, \SI{10}{\percent}, and \SI{10}{\percent}, respectively, using a stratified split by domain size. This stratification ensured that each domain size was proportionally represented across all three subsets, ensuring unbiased learning and evaluation across different microstructural scales.

\begin{table}[h!]
\centering
\footnotesize
\renewcommand{\arraystretch}{1.2}
\setlength{\extrarowheight}{2pt}
\begin{tabularx}{\textwidth}{p{2.2cm}>{\centering\arraybackslash}p{2.5cm}X}
\toprule
\textbf{Parameter} & \textbf{Value(s)} & \textbf{Description} \\
\midrule
$a$ (\si{\milli\meter}) & \num{2}, \num{3}, \num{4}, \num{5} & Side length of the square domain. \\ \cline{1-3}
Distribution & $\mathcal{N}(\overline{R},\,\sigma^{2})$ & The type of distribution used for the ECR distribution (Laguerre-Voronoï tessellation \cite{BERNACKI2024101224,Hitti2012}). \\ \cline{1-3}
$\overline{R}$ $(\SI{}{\micro\meter})$ & \num{20} & The mean value of the ECR distribution. \\ \cline{1-3}
$\sigma$ $(\SI{}{\micro\meter})$ & \num{2}, \num{4}, \num{8}, \num{16}, \num{32} & The standard deviation of the ECR distribution. \\
\bottomrule
\end{tabularx}
\caption{Parameters for generating initial microstructure states using LavoGen. Equivalent Circle Radius (ECR) = the radius of the circle having the same area as the considered grain (in \SI{}{\milli\meter} or \SI{}{\micro\meter}).}
\label{tab:dataset_gen_parameters_lavogen}
\end{table}

\begin{table}[h!]
\centering
\footnotesize
\renewcommand{\arraystretch}{1.2}
\setlength{\extrarowheight}{2pt}
\begin{tabularx}{\textwidth}{p{5cm}>{\centering\arraybackslash}p{2.5cm}X}
\toprule
\textbf{Parameter} & \textbf{Value(s)} & \textbf{Description} \\
\midrule
$T$ $(\SI{}{\kelvin})$ & \numrange{293.15}{1325.15} & The temperature range. \\ \cline{1-3}
$\mathrm{d}T/\mathrm{d}t$ heating $(\SI{}{\kelvin\per\second})$ & \numrange{0.1}{1} & The heating rate range. \\ \cline{1-3}
$|\mathrm{d}T/\mathrm{d}t|$ cooling $(\SI{}{\kelvin\per\second})$ & \numrange{0.01}{10} & The cooling rate range (absolute value). \\ \cline{1-3}
$Q$ $(\SI{}{\joule\per\mole})$ & \num{280000} & The activation energy. \\ \cline{1-3}
$\gamma$ $(\SI{}{\joule\per\meter\squared})$ & \num{0.6} & The grain boundary energy. \\ \cline{1-3}
$\mu_0$ $(\SI{}{\meter\tothe{4}\per\joule\per\second})$& \num{0.156} & The pre-exponential constant parameter. \\ \cline{1-3}
Simulation duration ($\SI{}{\minute}$) & \num{60} & The physical time of the simulation. \\
\bottomrule
\end{tabularx}
\caption{Physical parameters for simulating grain growth using TRM.}
\label{tab:dataset_gen_parameters_trm}
\end{table}

\subsection{Experimental Setup}
\label{sec:experimental_setup}

\begin{figure}[h!]
    \centering
    \begin{subfigure}[t]{0.48\linewidth}
    \centering
    \includegraphics[width=\linewidth]{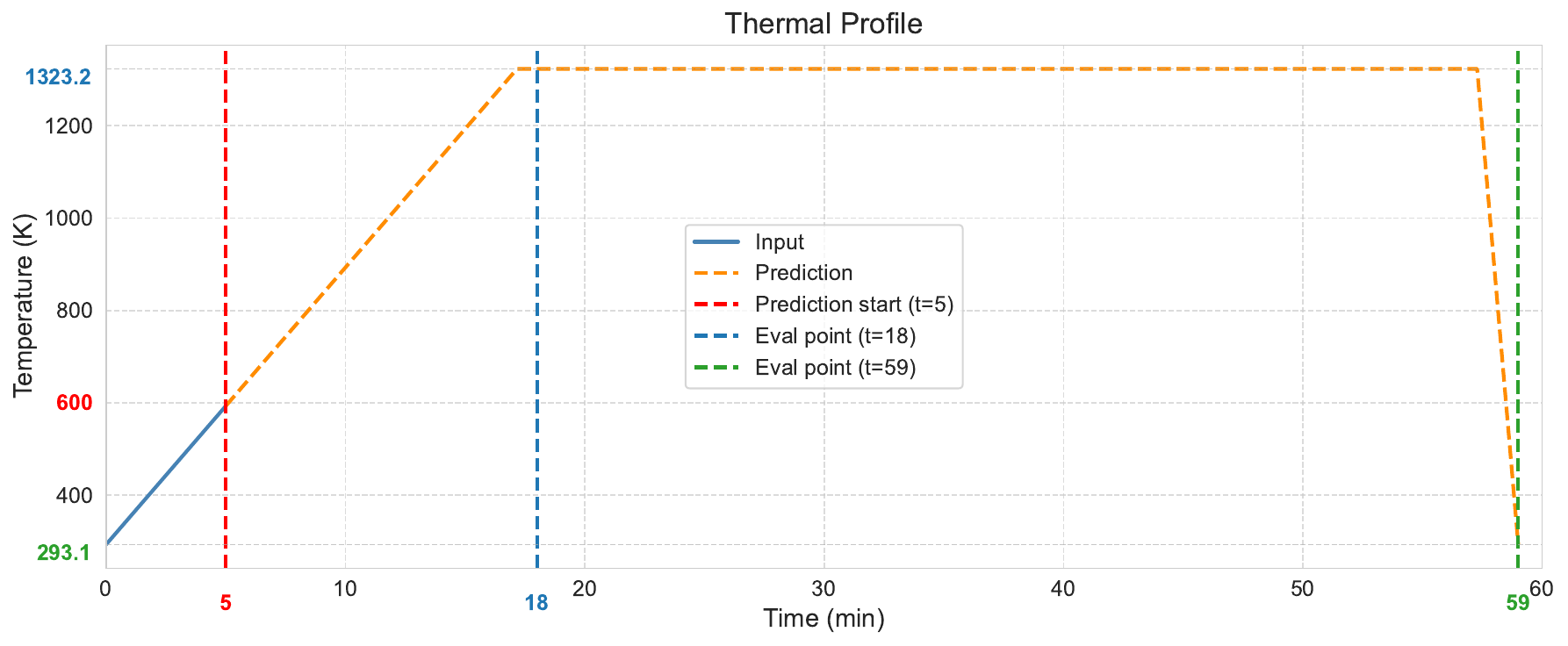}
    \caption{}
    \label{fig:thermal_profiles_s1_T}
    \end{subfigure}
    \hfill
    \begin{subfigure}[t]{0.48\linewidth}
    \centering
    \includegraphics[width=\linewidth]{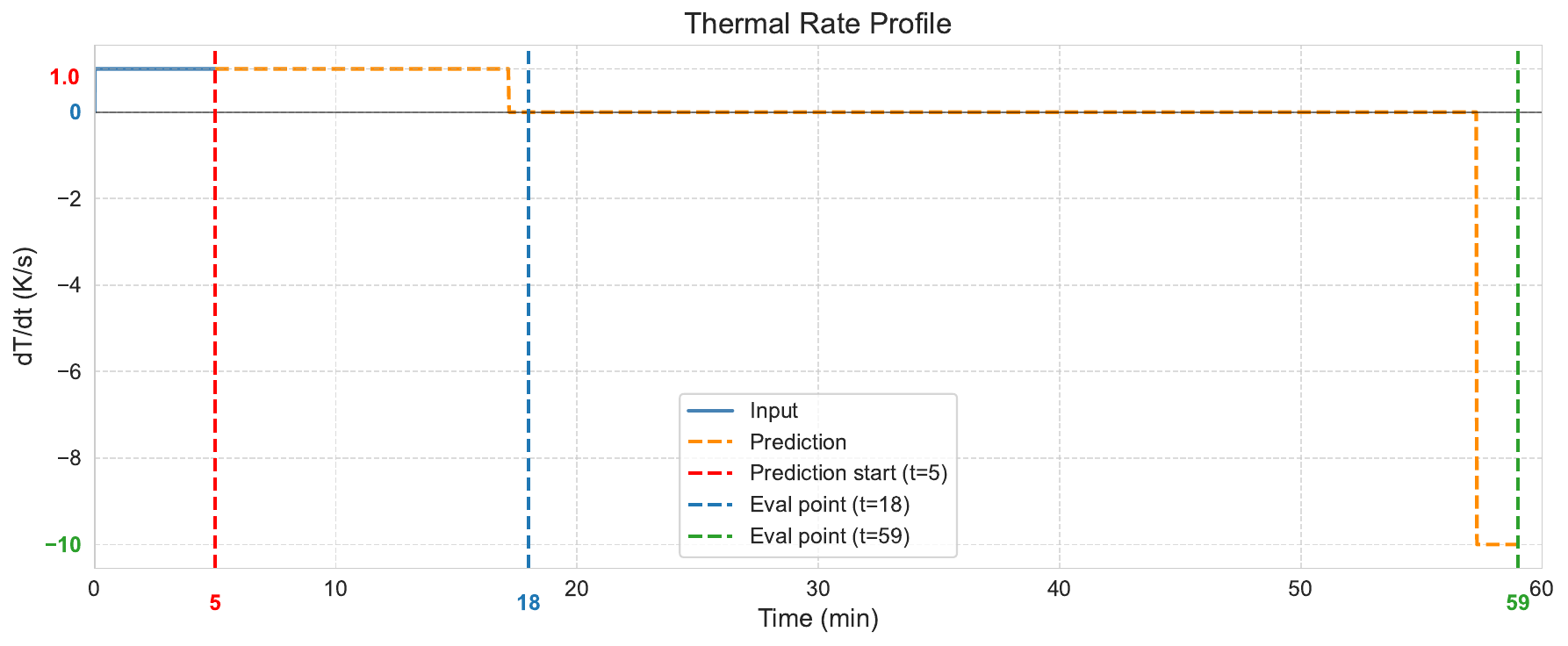}
    \caption{}
    \label{fig:thermal_profiles_s1_dT}
    \end{subfigure}

    \vspace{0.5em}

    \begin{subfigure}[t]{0.48\linewidth}
    \centering
    \includegraphics[width=\linewidth]{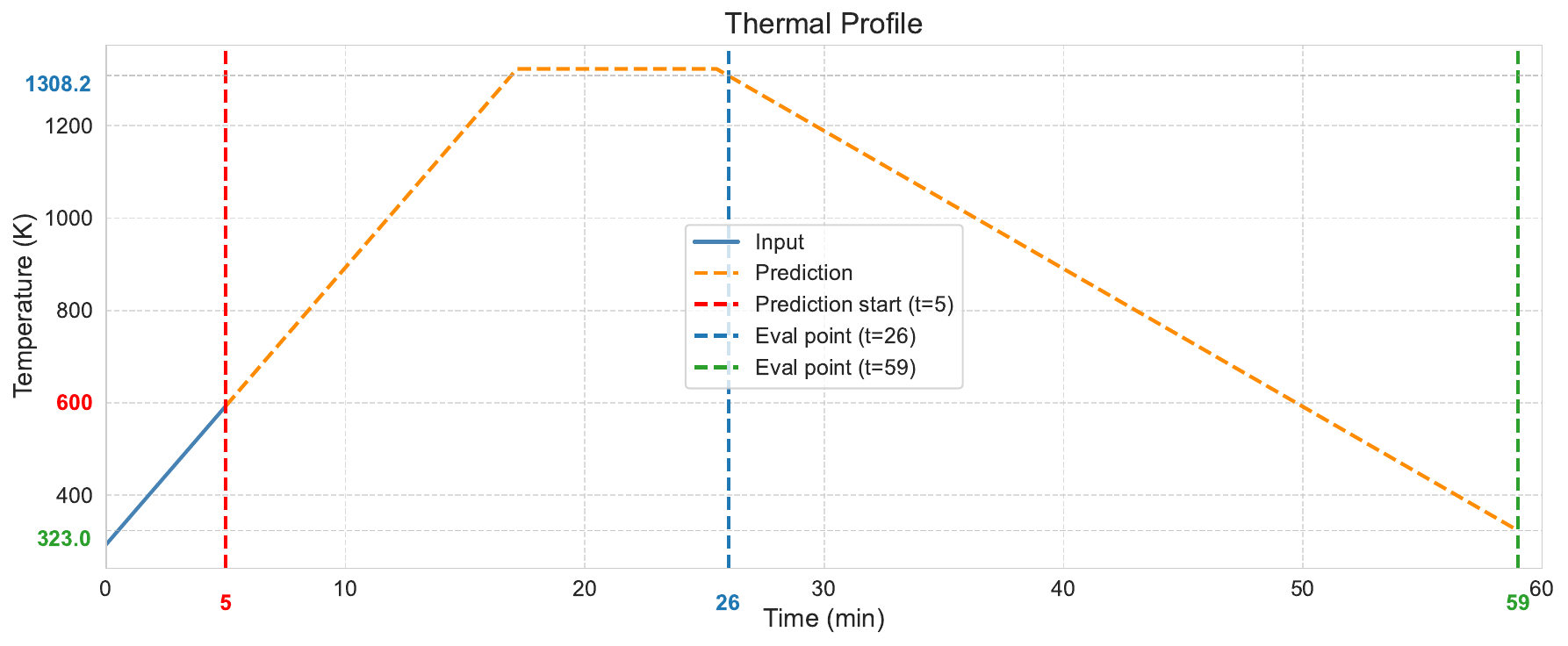}
    \caption{}
    \label{fig:thermal_profiles_s2_T}
    \end{subfigure}
    \hfill
    \begin{subfigure}[t]{0.48\linewidth}
    \centering
    \includegraphics[width=\linewidth]{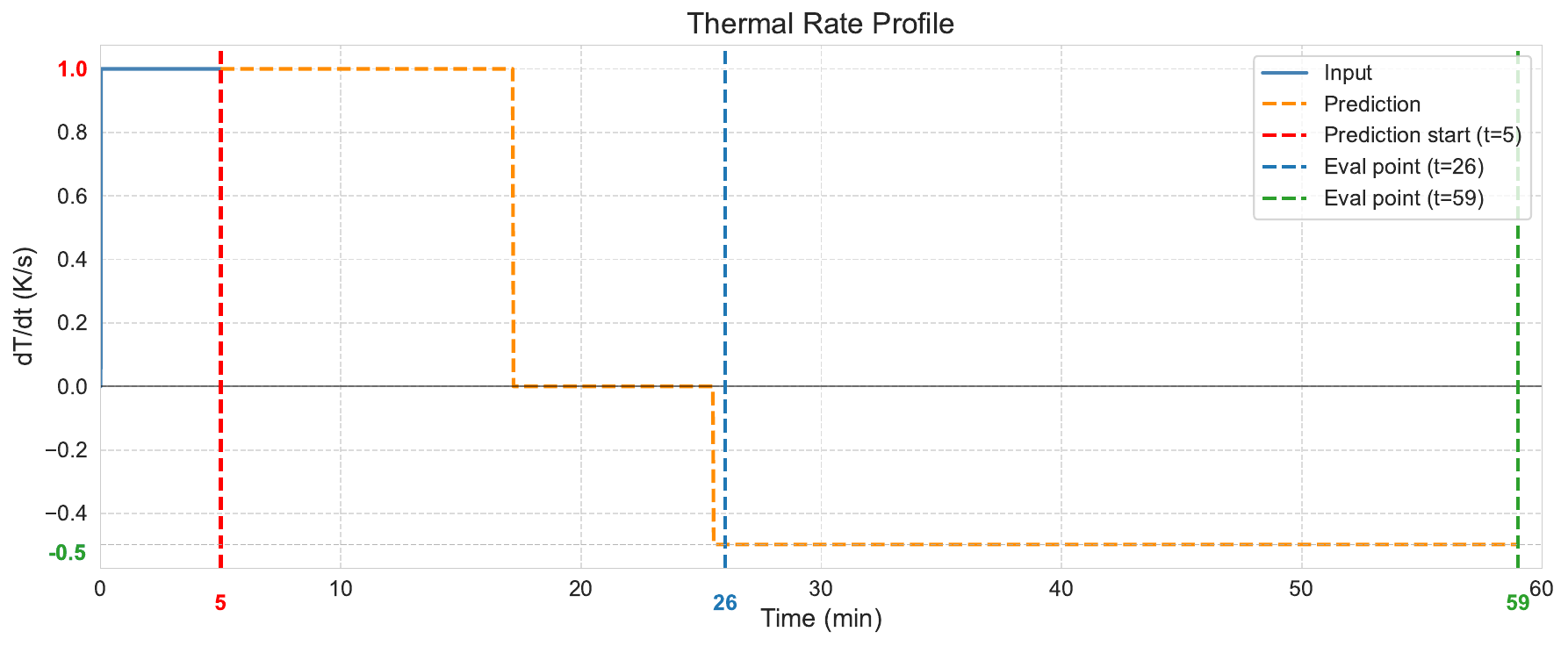}
    \caption{}
    \label{fig:thermal_profiles_s2_dT}
    \end{subfigure}

    \vspace{0.5em}

    \begin{subfigure}[t]{0.48\linewidth}
    \centering
    \includegraphics[width=\linewidth]{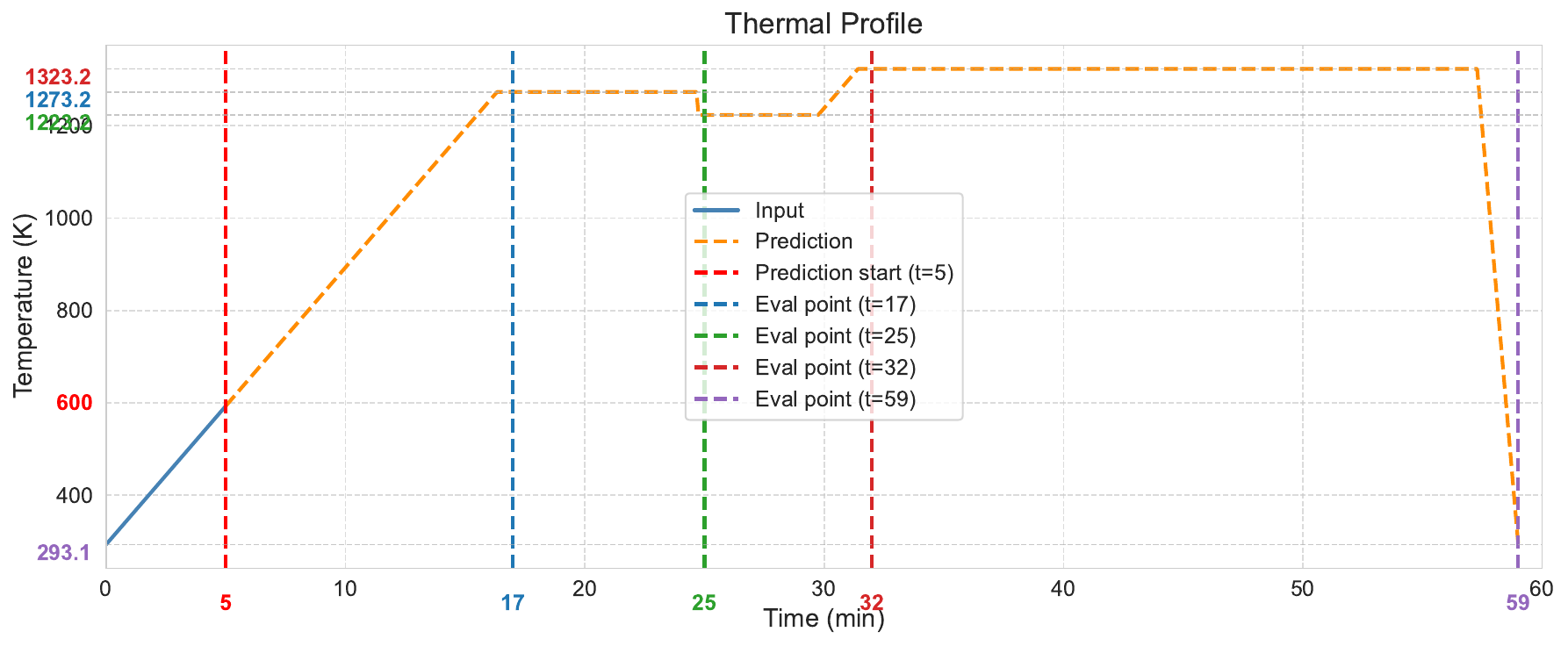}
    \caption{}
    \label{fig:thermal_profiles_s3_T}
    \end{subfigure}
    \hfill
    \begin{subfigure}[t]{0.48\linewidth}
    \centering
    \includegraphics[width=\linewidth]{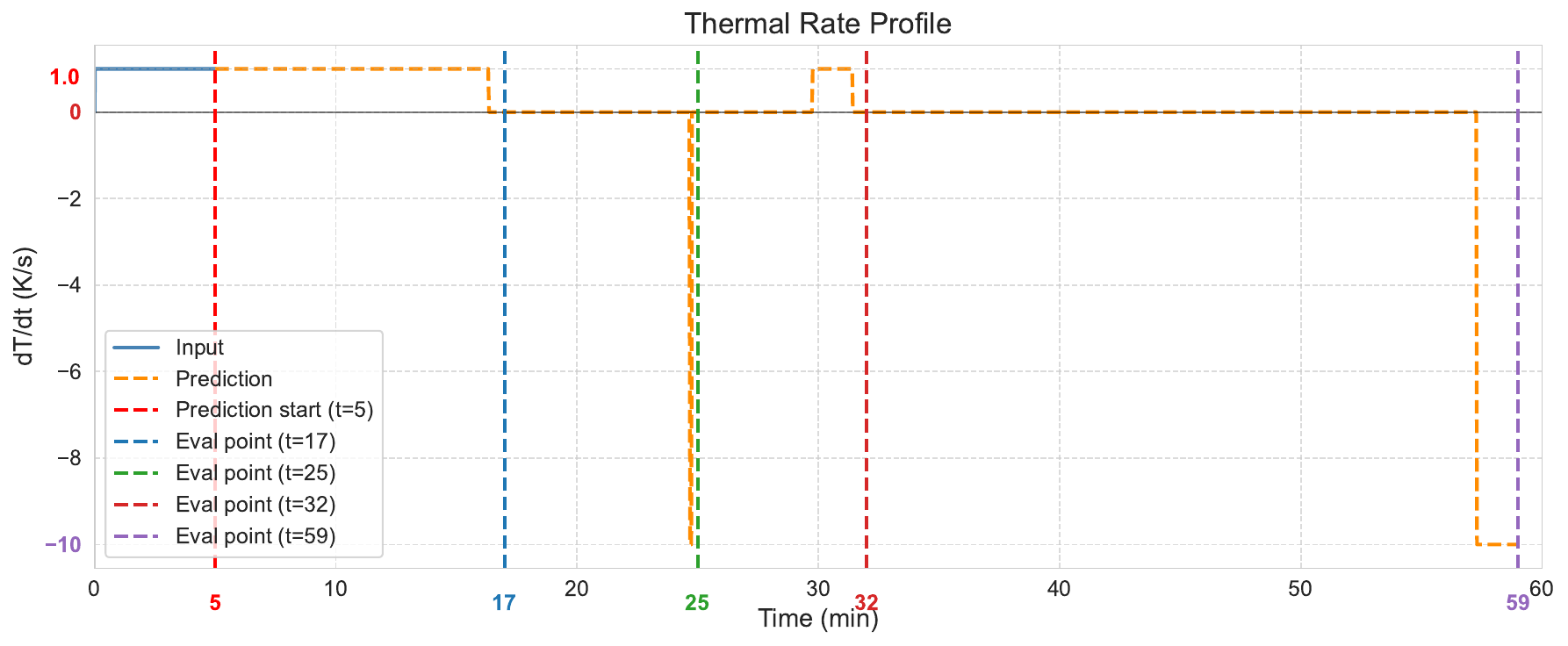}
    \caption{}
    \label{fig:thermal_profiles_s3_dT}
    \end{subfigure}

    \caption{Thermal profiles for the three test scenarios: temperature evolution (left column) and corresponding thermal rate $\mathrm{d}T/\mathrm{d}t$ (right column) for Scenario 1 (a, b), Scenario 2 (c, d), and Scenario 3 (e, f).}
    \label{fig:thermal_profiles}
\end{figure}

Three test scenarios of increasing complexity were designed to evaluate the capability of the model across diverse thermal conditions, as illustrated in Figure~\ref{fig:thermal_profiles}. Scenario 1 represents a complete heating-holding-cooling cycle consisting of heating from \SI{293.15}{\kelvin} to \SI{1323.15}{\kelvin} at \SI{1}{\kelvin\per\second}, an approximately \SI{39}{\minute} hold, and rapid cooling at \SI{10}{\kelvin\per\second}. Scenario 2 examines slow cooling at \SI{0.5}{\kelvin\per\second} following a shorter holding period, isolating the error behavior during the cooling phase independently of prior error accumulation. Scenario 3 presents a complex multi-cycle profile intentionally excluded from training, designed to provide a rigorous test of generalization to novel thermal histories. This profile consists of sequential heating, partial cooling, reheating, extended holding, and rapid quenching stages. Reference microstructure states simulated with the TRM for Scenarios 1, 2, and 3 are provided in Supplementary Figures~\ref{fig:s1_scenario1}, \ref{fig:s1_scenario2}, and \ref{fig:s1_scenario3}, respectively.

For all scenarios, the model was provided the first \SI{5}{\minute} of evolution as input and generated predictions autoregressively in 5-minute segments up to $t = \SI{59}{\minute}$. Performance was evaluated at critical time points corresponding to the end of each thermal phase of interest.

\subsection{Evaluation Metrics}
\label{sec:metrics}

Prediction performance was assessed across three complementary categories, extending the evaluation framework of the previous study with one additional metric. Pixel-wise accuracy was quantified using Boundary-Focused Mean Absolute Error ($\text{MAE}_{\text{b}}$)~\cite{TEP2025121486}, Boundary-Focused Mean Squared Error ($\text{MSE}_{\text{b}}$)~\cite{TEP2025121486}, Peak Signal-to-Noise Ratio (PSNR)~\cite{hore2010image}, and SSIM~\cite{hore2010image}. Grain-level statistical fidelity was assessed through Kullback-Leibler (KL) divergence~\cite{kullback1951information}, Wasserstein distance~\cite{villani2008optimal}, and $\overline{R}$ error computed on the surface-weighted grain size distribution characterized by  Equivalent Circle Radius (ECR) extracted through image processing on predicted images. A grain-topology metric was additionally introduced, namely the distribution of the number of neighbors per grain, which characterizes grain network connectivity and verifies that predictions maintain topological characteristics consistent with two-dimensional grain growth, where approximately six neighbors per grain is expected under curvature-driven boundary migration~\cite{mullins1956two}.


\section{Results}
\label{sec:results}

\subsection{Scenario 1: Heating-Holding-Cooling Cycle}

\begin{figure}[h!]
    \centering
    \begin{subfigure}[t]{0.48\linewidth}
    \centering
    \includegraphics[width=\linewidth]{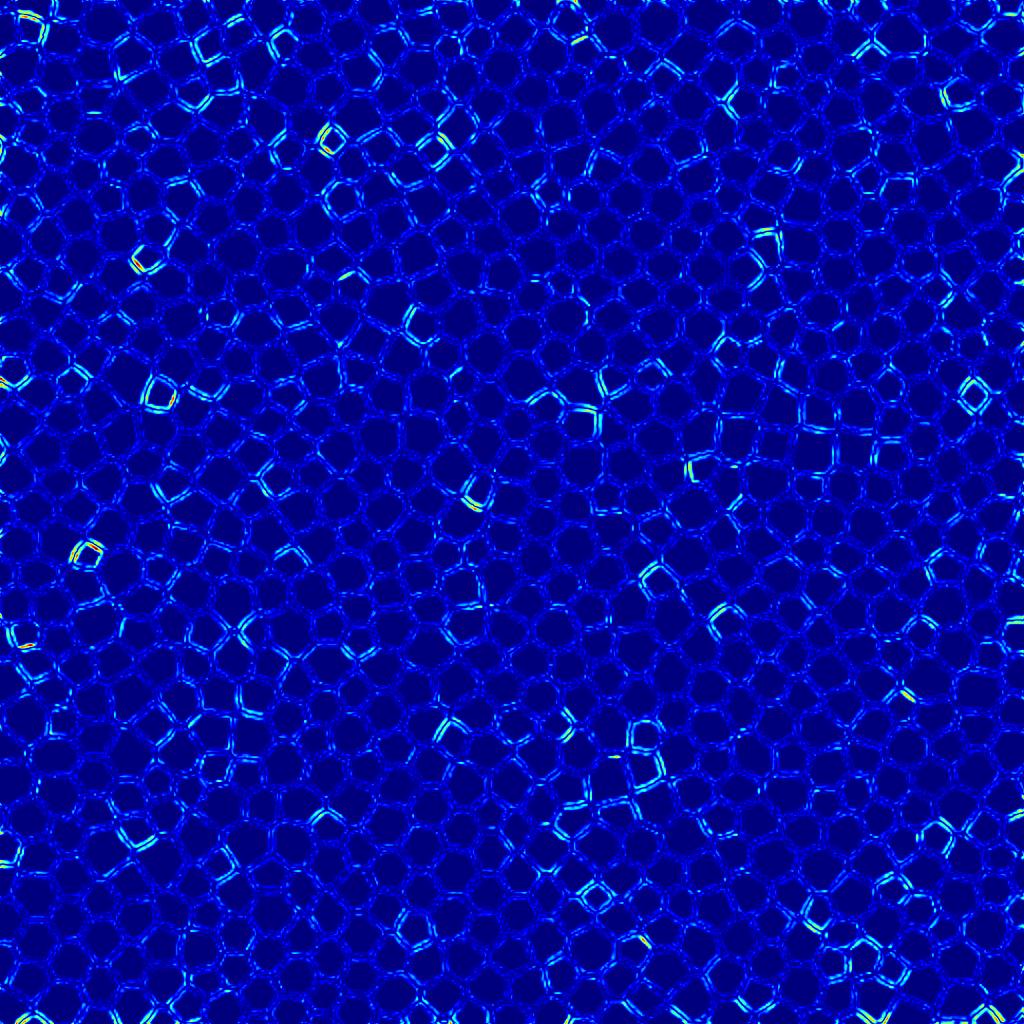}
    \caption{}
    \label{fig:result_scenario_1_a}
    \end{subfigure}
    \hfill
    \begin{subfigure}[t]{0.48\linewidth}
    \centering
    \includegraphics[width=\linewidth]{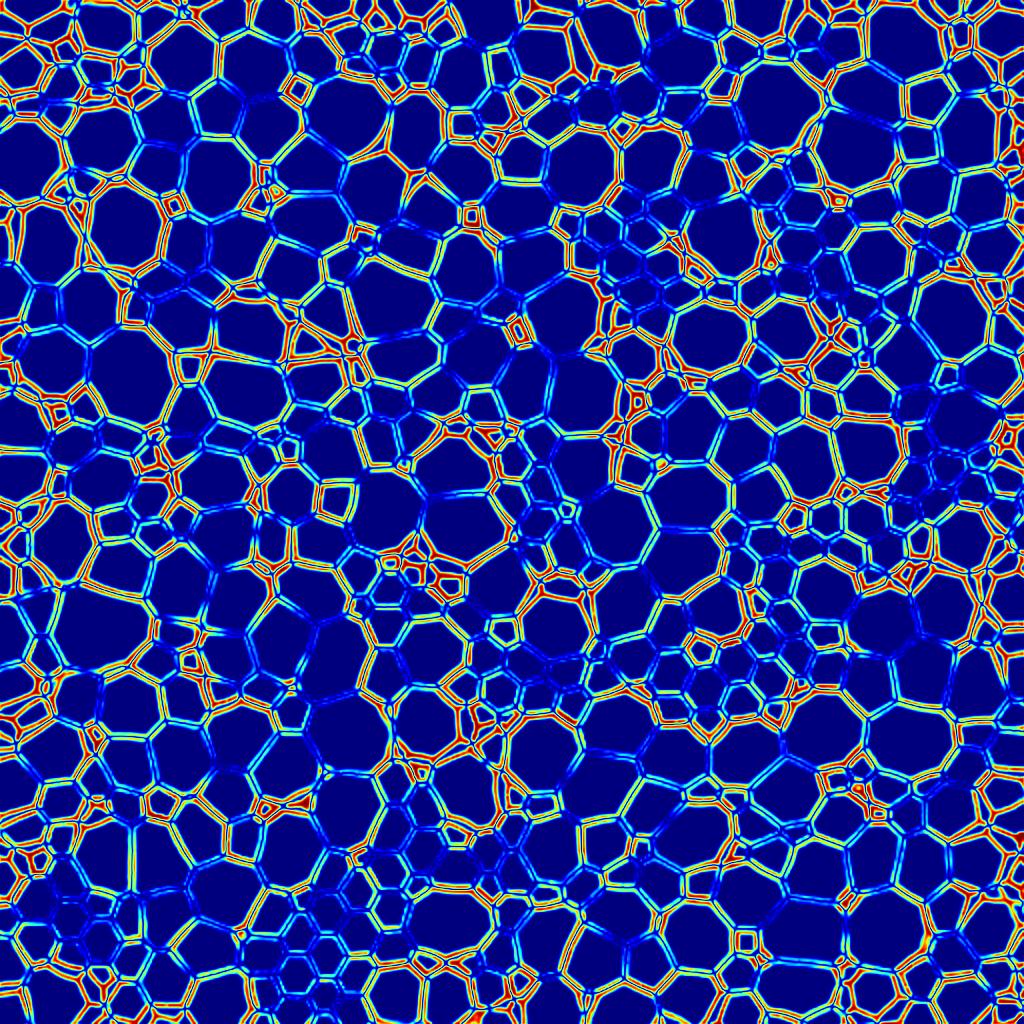}
    \caption{}
    \label{fig:result_scenario_1_b}
    \end{subfigure}
    \vspace{0.10cm}
    \centering
    \includegraphics[width=1\textwidth]{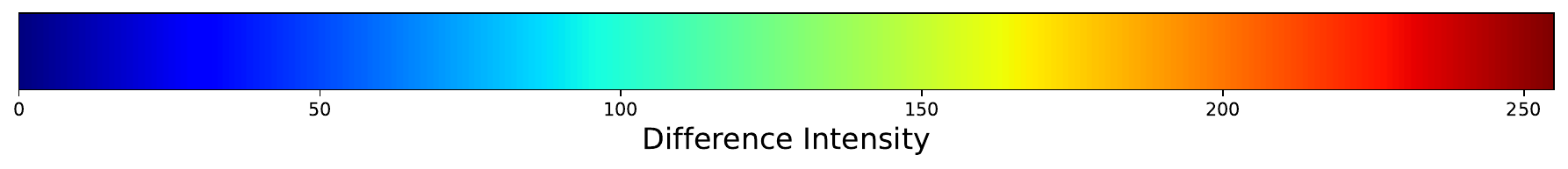}
    \caption{Error heatmap for Scenario 1: (a) end of heating phase ($t = \SI{18}{\minute}$); (b) end of cooling phase ($t = \SI{59}{\minute}$). (For interpretation of the references to color in this figure legend, the reader is referred to the web version of this article.)}
    \label{fig:result_scenario_1}

\end{figure}

\begin{table}[htbp]
    \centering

    \resizebox{\textwidth}{!}{%
    \begin{tabular}{c|cccc|ccc|w{c}{4em}w{c}{4em}}
    \toprule
    \multirow{2}{*}{\begin{tabular}{c} \textbf{Evaluation} \\ \textbf{Time Point ($t$, min)} \end{tabular}} & \multicolumn{4}{c|}{\textbf{Pixel-wise and Perceptual}} & \multicolumn{3}{c|}{\textbf{ECR Distribution}} & \multicolumn{2}{c}{\textbf{Neighbor Count Distribution}} \\
    \cmidrule(lr){2-5} \cmidrule(lr){6-8} \cmidrule(lr){9-10}
    & $\text{MAE}_\text{b}$ $\downarrow$ & $\text{MSE}_\text{b}$ $\downarrow$ & PSNR (dB) $\uparrow$ & SSIM $\uparrow$ & $\overline{R}$ Err. (\%) $\downarrow$ & KL $\downarrow$ & W $\downarrow$ & KL $\downarrow$ & W $\downarrow$ \\
    \midrule
    18 & 0.1629 & 0.0453 & 17.45 & 0.8384 & 0.035 & 0.0034 & 0.000150 & 0.0006 & 0.0168 \\
    59 & 0.2756 & 0.1148 & 10.45 & 0.5426 & 10.17 & 0.1823 & 0.006986 & 0.0547 & 0.1619 \\
    \bottomrule
    \end{tabular}%
    }

    \vspace{2pt}
    {\footnotesize KL = KL divergence (predicted $\rightarrow$ ground truth); W = Wasserstein distance; $\overline{R}$ = mean grain size (surface-weighted).\par}
    \caption{Quantitative evaluation metrics for Scenario 1 (heating-holding-cooling cycle).}
    \label{tab:scenario1_metrics}
\end{table}

The simple heating-holding-cooling cycle established a baseline for thermal conditioning performance. Quantitative metrics are summarized in Table~\ref{tab:scenario1_metrics}. At the end of the heating phase ($t = \SI{18}{\minute}$), the model achieved an SSIM of \num{0.8384} and $\overline{R}$ error of \SI{0.035}{\percent}, indicating accurate capture of boundary migration kinetics during the heating phase. The grain neighbor count distribution showed close alignment with the reference data (KL divergence of \num{6e-4}), reproducing the characteristic distribution centered around six neighbors per grain consistent with two-dimensional grain growth. At the end of the evolution ($t = \SI{59}{\minute}$), pixel-wise metrics showed more pronounced degradation, with the SSIM declining to \num{0.5426} and $\overline{R}$ error rising to \SI{10.17}{\percent}. The error heatmaps in Figure~\ref{fig:result_scenario_1} confirmed that discrepancies were concentrated in regions containing small grains approaching disappearance and complex junctions, while the overall grain boundary topology remained physically consistent. The grain neighbor count KL divergence increased from \num{0.06e-2} to \num{5.47e-2} between the two evaluation points, consistent with the pixel-wise degradation trend. ECR and neighbor count distributions are provided in Supplementary Figure~\ref{fig:s2_dist_scenario1}.

\subsection{Scenario 2: Slow Cooling Conditions}

\begin{figure}[h!]
    \centering
    \begin{subfigure}[t]{0.48\linewidth}
    \centering
    \includegraphics[width=\linewidth]{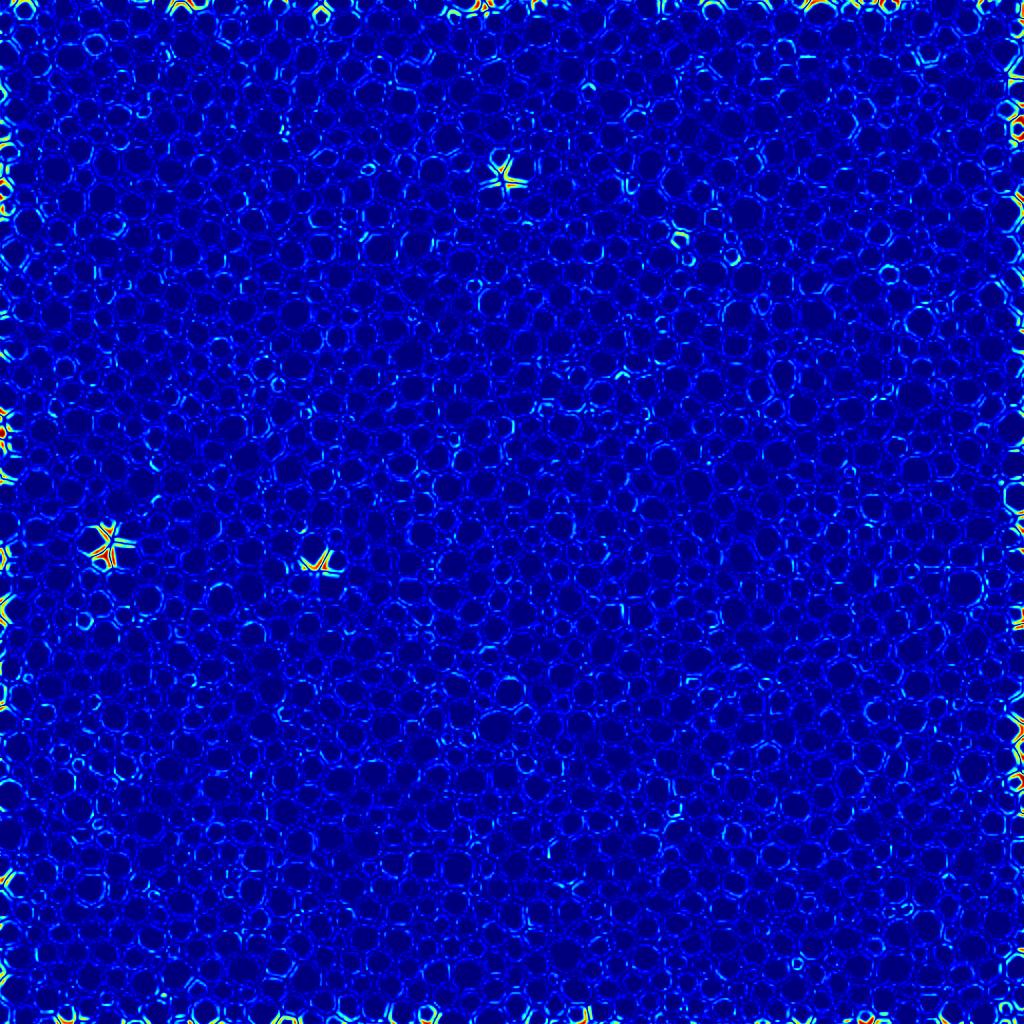}
    \caption{}
    \label{fig:result_scenario_2_a}
    \end{subfigure}
    \hfill
    \begin{subfigure}[t]{0.48\linewidth}
    \centering
    \includegraphics[width=\linewidth]{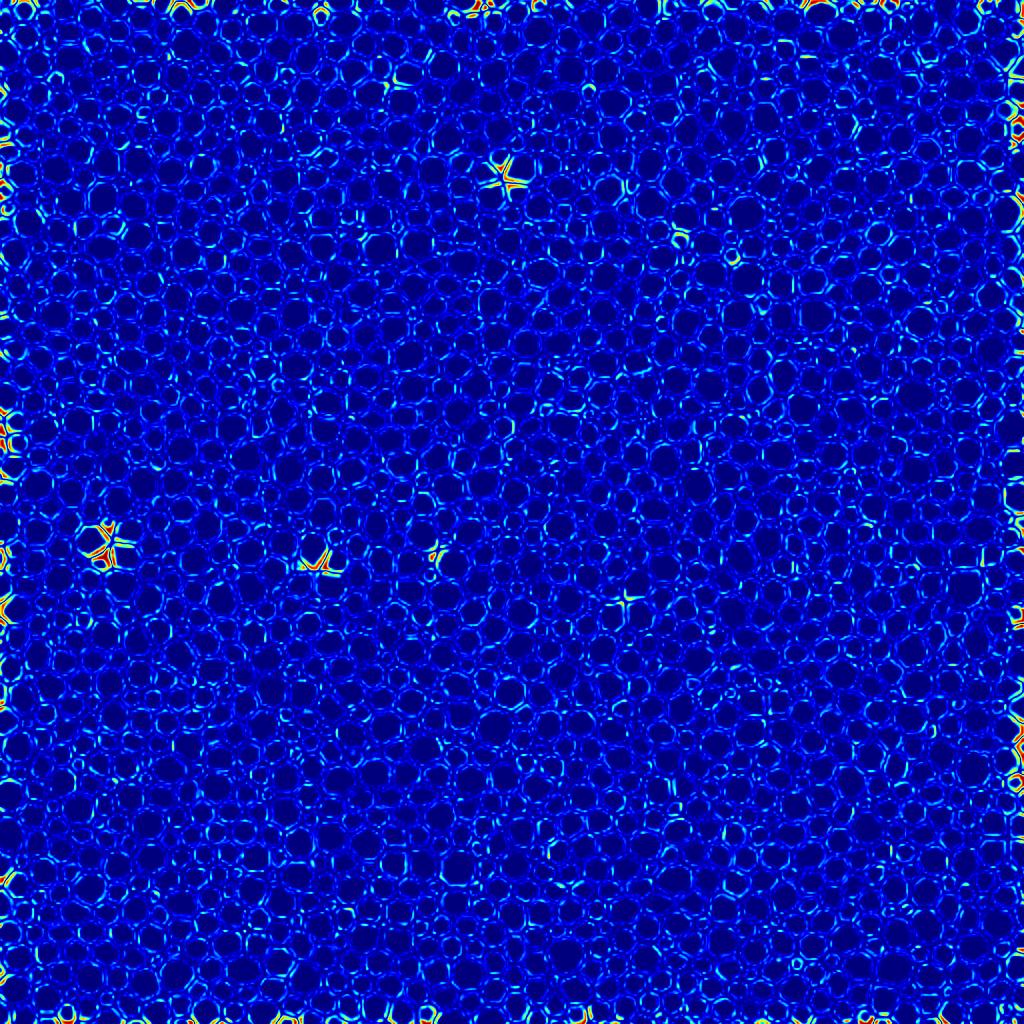}
    \caption{}
    \label{fig:result_scenario_2_b}
    \end{subfigure}
    \vspace{0.10cm}
    \centering
    \includegraphics[width=1\textwidth]{figures/heatmap_legend.pdf}
    \caption{Error heatmap for Scenario 2: (a) start of cooling phase ($t = \SI{26}{\minute}$); (b) end of cooling phase ($t = \SI{59}{\minute}$). (For interpretation of the references to color in this figure legend, the reader is referred to the web version of this article.)}
    \label{fig:result_scenario_2}
\end{figure}

\begin{table}[htbp]
    \centering

    \resizebox{\textwidth}{!}{%
    \begin{tabular}{c|cccc|ccc|w{c}{4em}w{c}{4em}}
    \toprule
    \multirow{2}{*}{\begin{tabular}{c} \textbf{Evaluation} \\ \textbf{Time Point ($t$, min)} \end{tabular}} & \multicolumn{4}{c|}{\textbf{Pixel-wise and Perceptual}} & \multicolumn{3}{c|}{\textbf{ECR Distribution}} & \multicolumn{2}{c}{\textbf{Neighbor Count Distribution}} \\
    \cmidrule(lr){2-5} \cmidrule(lr){6-8} \cmidrule(lr){9-10}
    & $\text{MAE}_\text{b}$ $\downarrow$ & $\text{MSE}_\text{b}$ $\downarrow$ & PSNR (dB) $\uparrow$ & SSIM $\uparrow$ & $\overline{R}$ Err. (\%) $\downarrow$ & KL $\downarrow$ & W $\downarrow$ & KL $\downarrow$ & W $\downarrow$ \\
    \midrule
    26 & 0.1105 & 0.0213 & 20.18 & 0.9172 & 0.66 & 0.0042 & 0.000193 & 0.0032 & 0.0672 \\
    59 & 0.1325 & 0.0294 & 18.93 & 0.8873 & 0.11 & 0.0047 & 0.000124 & 0.0086 & 0.0134 \\
    \bottomrule
    \end{tabular}%
    }

    \vspace{2pt}
    {\footnotesize KL = KL divergence (predicted $\rightarrow$ ground truth); W = Wasserstein distance; $\overline{R}$ = mean grain size (surface-weighted).\par}

    \caption{Quantitative evaluation metrics for Scenario 2 (slow cooling at \SI{0.5}{\kelvin\per\second}).}
    \label{tab:scenario2_metrics}
\end{table}

In contrast, the slow cooling scenario (\SI{0.5}{\kelvin\per\second}) was designed to isolate the error behavior during the cooling phase found in Scenario 1 by reducing error accumulation through a shorter holding period. At the start of the cooling phase ($t = \SI{26}{\minute}$), the model achieved an SSIM of \num{0.9172} and $\overline{R}$ error of \SI{0.66}{\percent}. Over the subsequent \SI{33}{\minute} slow cooling phase, metrics remained stable, with the SSIM decreasing by only \SI{3.3}{\percent} to \num{0.8873}, $\overline{R}$ error improving to \SI{0.11}{\percent}, and the KL divergence for the ECR distribution increasing only marginally from \num{4.2e-3} to \num{4.7e-3}. The grain neighbor count distributions maintained close agreement with the reference data throughout the cooling phase, with a KL divergence of \num{3.2e-3} at the start and \num{8.6e-3} at the end of cooling, as confirmed by the error heatmaps in Figure~\ref{fig:result_scenario_2}. ECR and neighbor count distributions for this scenario are provided in Supplementary Figure~\ref{fig:s2_dist_scenario2}. Quantitative metrics are summarized in Table~\ref{tab:scenario2_metrics}.

\subsection{Scenario 3: Complex Multi-Cycle Thermal Profile}

\begin{figure}[h!]
    \centering
    \begin{subfigure}[t]{0.48\linewidth}
    \centering
    \includegraphics[width=\linewidth]{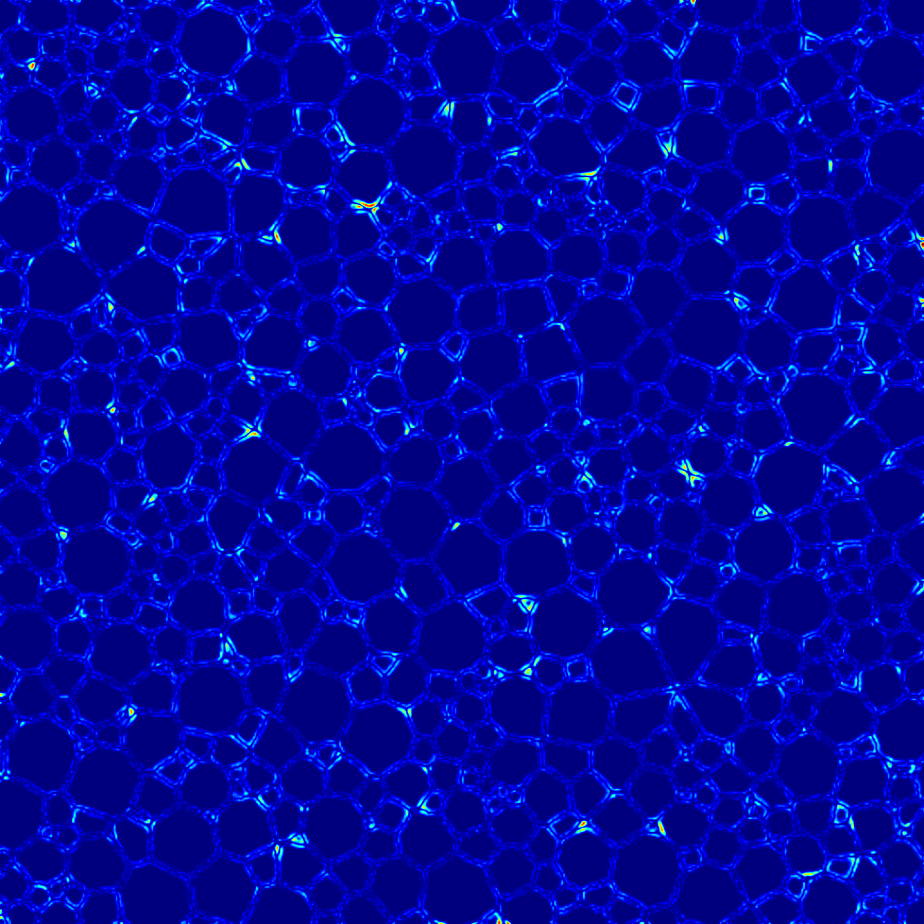}
    \caption{}
    \label{fig:result_scenario_3_a}
    \end{subfigure}
    \hfill
    \begin{subfigure}[t]{0.48\linewidth}
    \centering
    \includegraphics[width=\linewidth]{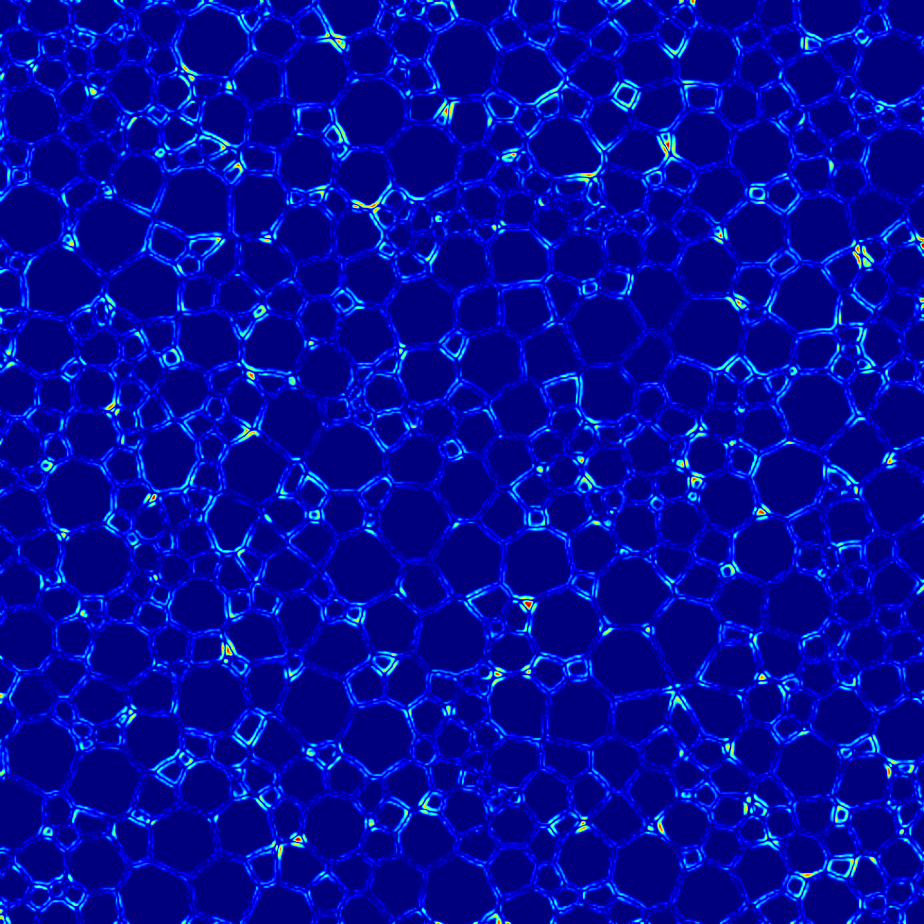}
    \caption{}
    \label{fig:result_scenario_3_b}
    \end{subfigure}

    \vspace{0.75em}

    \begin{subfigure}[t]{0.48\linewidth}
    \centering
    \includegraphics[width=\linewidth]{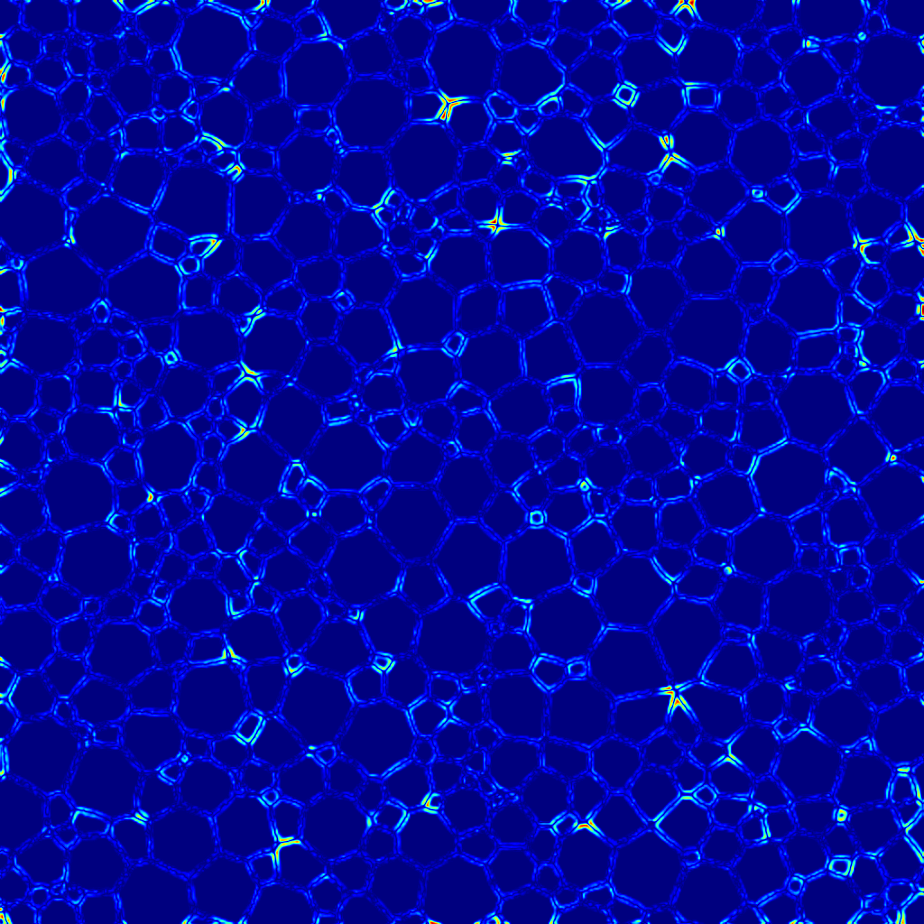}
    \caption{}
    \label{fig:result_scenario_3_c}
    \end{subfigure}
    \hfill
    \begin{subfigure}[t]{0.48\linewidth}
    \centering
    \includegraphics[width=\linewidth]{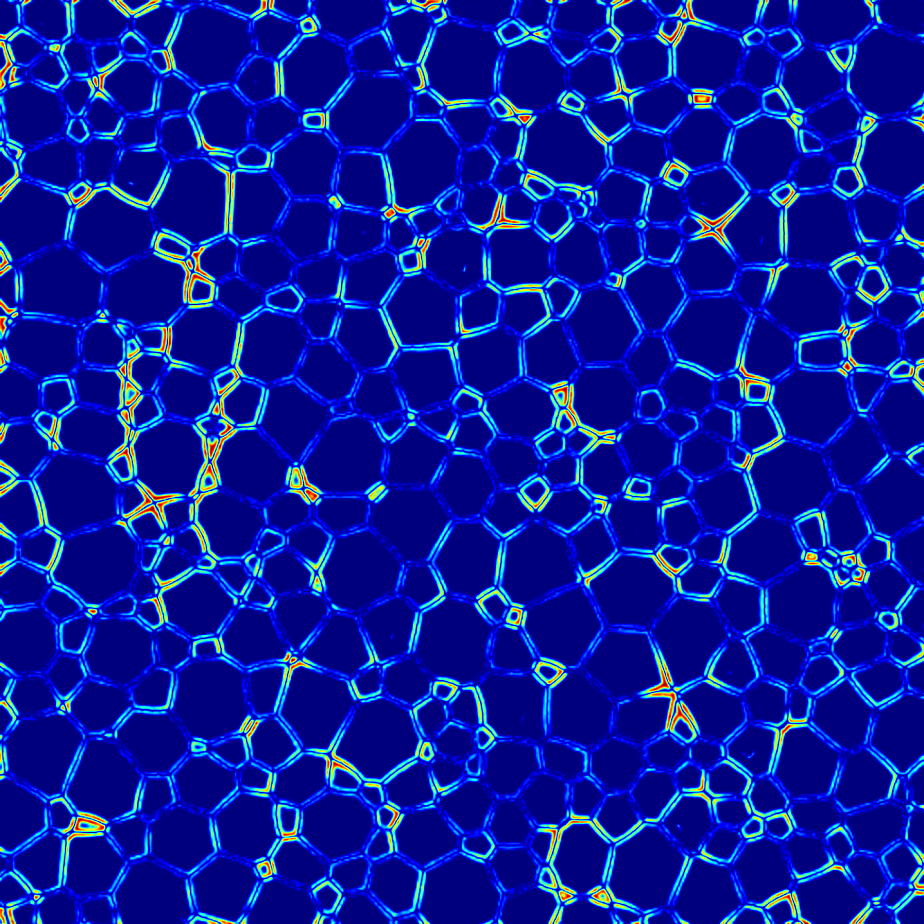}
    \caption{}
    \label{fig:result_scenario_3_d}
    \end{subfigure}
    \vspace{0.10cm}
    \centering
    \includegraphics[width=1\textwidth]{figures/heatmap_legend.pdf}
    \caption{Error heatmap for Scenario 3: (a) end of first heating ($t = \SI{17}{\minute}$); (b) end of partial cooling ($t = \SI{25}{\minute}$); (c) end of reheating ($t = \SI{32}{\minute}$); (d) end of final cooling ($t = \SI{59}{\minute}$). (For interpretation of the references to color in this figure legend, the reader is referred to the web version of this article.)}
    \label{fig:result_scenario_3}
\end{figure}

\begin{table}[htbp]
    \centering

    \resizebox{\textwidth}{!}{%
    \begin{tabular}{c|cccc|ccc|w{c}{4em}w{c}{4em}}
    \toprule
    \multirow{2}{*}{\begin{tabular}{c} \textbf{Evaluation} \\ \textbf{Time Point ($t$, min)} \end{tabular}} & \multicolumn{4}{c|}{\textbf{Pixel-wise and Perceptual}} & \multicolumn{3}{c|}{\textbf{ECR Distribution}} & \multicolumn{2}{c}{\textbf{Neighbor Count Distribution}} \\
    \cmidrule(lr){2-5} \cmidrule(lr){6-8} \cmidrule(lr){9-10}
    & $\text{MAE}_\text{b}$ $\downarrow$ & $\text{MSE}_\text{b}$ $\downarrow$ & PSNR (dB) $\uparrow$ & SSIM $\uparrow$ & $\overline{R}$ Err. (\%) $\downarrow$ & KL $\downarrow$ & W $\downarrow$ & KL $\downarrow$ & W $\downarrow$ \\
    \midrule
    17 & 0.1138 & 0.0225 & 22.43 & 0.9264 & 1.05 & 0.0026 & 0.000579 & 0.0018 & 0.0320 \\
    25 & 0.1160 & 0.0237 & 22.29 & 0.9278 & 2.09 & 0.0105 & 0.001269 & 0.0080 & 0.0778 \\
    32 & 0.1214 & 0.0263 & 21.73 & 0.9240 & 2.57 & 0.0127 & 0.001549 & 0.0085 & 0.0600 \\
    59 & 0.2045 & 0.0674 & 15.77 & 0.7834 & 3.20 & 0.0087 & 0.002161 & 0.0468 & 0.1630 \\
    \bottomrule
    \end{tabular}%
    }

    \vspace{2pt}
    {\footnotesize KL = KL divergence (predicted $\rightarrow$ ground truth); W = Wasserstein distance; $\overline{R}$ = mean grain size (surface-weighted).\par}

    \caption{Quantitative evaluation metrics for Scenario 3 (complex multi-cycle thermal profile, excluded from training data).}
    \label{tab:scenario3_metrics}
\end{table}

Finally, the complex multi-cycle profile, excluded from the training data, provided the most rigorous test of the generalization capability of the model. Through the first three evaluation points spanning multiple heating-cooling transitions ($t = \SI{17}{\minute}$, $\SI{25}{\minute}$, $\SI{32}{\minute}$), the SSIM remained stable between \num{0.9240} and \num{0.9278}, and $\overline{R}$ error remained below \SI{3}{\percent}. The grain neighbor count KL divergence remained below \num{8.5e-3} across these evaluation points, indicating that the predicted microstructures maintained topological consistency with the reference data despite the novel thermal history. At the final evaluation point following the extended isothermal hold and rapid cooling ($t = \SI{59}{\minute}$), the SSIM declined to \num{0.7834} and $\overline{R}$ error reached \SI{3.20}{\percent}, consistent with the error accumulation pattern observed in Scenario 1. Despite this degradation, the error heatmaps in Figure~\ref{fig:result_scenario_3} confirmed that the overall grain boundary topology remained physically consistent, with deviations concentrated at small grains and complex junctions. ECR and neighbor count distributions across all evaluation points are provided in Supplementary Figure~\ref{fig:s2_dist_scenario3}. Quantitative metrics are summarized in Table~\ref{tab:scenario3_metrics}.


\section{Discussion}
\label{sec:discussion}

Beyond pixel-level correspondence, the error heatmap comparisons across all scenarios indicated that the overall grain boundary topology was preserved even at evaluation points where pixel-wise metrics showed more pronounced degradation. The predicted microstructures retained the essential features relevant to domain expert analysis, including the spatial arrangement of grain boundaries, ECR distributions, and neighbor count distributions. The ECR and neighbor count distributions provided additional quantitative support for this observation. In Scenarios 2 and 3, $\overline{R}$ error remained below \SI{3.2}{\percent} and KL divergence values for ECR distributions remained low throughout the prediction horizon. The neighbor count distributions reproduced the characteristic distribution centered around six neighbors per grain across all scenarios, with KL divergences remaining below \num{9e-3} at early evaluation points. Together, these distributional metrics indicate that the predicted microstructures capture the statistically relevant characteristics suitable for practical microstructural analysis, even in cases where pixel-wise metrics indicate moderate degradation.

Additionally, the temporal evolution of prediction errors revealed a consistent pattern across all scenarios, where error accumulation within individual thermal phases remained relatively modest while more pronounced degradation occurred specifically during extended isothermal holding phases. In Scenario 2, the SSIM decreased by only \SI{3.3}{\percent} over a \SI{33}{\minute} slow cooling phase, confirming that the thermal conditioning mechanism does not introduce substantial errors during active thermal phases. The more pronounced degradation at the end of Scenarios 1 and 3 occurred during extended isothermal holds at peak temperature, where prolonged grain coarsening amplified small prediction errors progressively. This behavior is consistent with the established characteristics of autoregressive sequence-to-sequence prediction and was similarly observed in the previous study under isothermal conditions.

Overall, the integration of thermal conditioning through FiLM demonstrated measurable benefits for grain growth prediction under complex thermal profiles, with the model adapting its predictions based on instantaneous $T$ and $\mathrm{d}T/\mathrm{d}t$ to reproduce grain boundary kinetics across heating, isothermal holding, and cooling phases. The comparison between Scenarios 1 and 2 provided direct evidence for the effectiveness of this mechanism, with the modest error increase over the slow cooling phase confirming that the learned modulation functions appropriately scaled grain boundary kinetics in response to thermal rate variations. The acceptable performance on Scenario 3, with $\overline{R}$ error below \SI{3.2}{\percent}, suggests that the model has learned generalizable thermal conditioning rules rather than memorizing specific thermal profile patterns from the training data. The integration of thermal conditioning modules does not compromise inference speed, which remains on the order of seconds per prediction sequence (approximately \SI{15}{\second}). Training duration increased from \SI{3}{\hour} to \SI{60}{\hour} due to both the substantially larger dataset (\num{6727} versus \num{648} sequences) required to ensure sufficient coverage of the thermal parameter space and the additional computational cost introduced by the FiLM thermal conditioning module at each training step.

Nevertheless, the primary limitation of the current framework is error accumulation during extended isothermal holding phases, where prolonged grain coarsening amplifies small boundary position errors over time. Addressing this may require modifications to the autoregressive prediction strategy, such as the periodic incorporation of intermediate reference states or the integration of physics-informed constraints to mitigate long-term error compounding.


\section{Conclusion}

This paper demonstrated the extension of the DL framework established in the previous study for grain growth prediction to complex thermal profiles through the integration of FiLM for thermal conditioning. By providing thermal descriptors $T$ and $\mathrm{d}T/\mathrm{d}t$ as explicit inputs that modulate latent feature representations, the model adapted its predictions according to instantaneous thermal conditions rather than relying on implicit spatiotemporal patterns learned from constant-rate training data. The architectural modifications required for thermal conditioning, namely the thermal conditioning module and FiLM layers, introduce minimal computational overhead, with inference time remaining on the order of seconds per prediction sequence. This preserves the primary advantage of the DL approach over traditional PDE-based simulations, which require substantially longer computation times due to their sequential calculation in each timestep.

Specifically, evaluation across three test scenarios of increasing complexity revealed several key findings. First, the thermally conditioned model achieved acceptable prediction accuracy across diverse thermal phases. In Scenarios 2 and 3, the SSIM remained above \num{0.78} and $\overline{R}$ error stayed below \SI{3.2}{\percent} at all evaluation points. Scenario 1, which involved an extended approximately \SI{39}{\minute} isothermal holding phase, exhibited more pronounced degradation at the final time point, with the decrease in accuracy attributed primarily to autoregressive error accumulation during the prolonged isothermal holding rather than to limitations of the thermal conditioning mechanism. Second, the ECR and neighbor count distributions extracted from predicted microstructures showed good alignment with ground truth, indicating that the model preserves statistically relevant microstructural characteristics suitable for practical analysis. Third, the performance of the model on a complex multi-cycle thermal profile excluded from training suggests that conditioning through FiLM captures generalizable relationships between thermal descriptors and microstructure rather than memorizing specific thermal patterns.

Furthermore, the results identified the conditions under which error accumulation is most pronounced. By comparing Scenarios 1 and 2, which evaluate distinct phases of the thermal cycle, it was demonstrated that error accumulation within individual heating or cooling phases remained modest, with the SSIM decreasing by only \SI{3.3}{\percent} over a \SI{33}{\minute} slow cooling phase in Scenario 2, confirming that the prolonged isothermal hold at peak temperature was the primary driver of the pronounced degradation observed at the end of Scenario 1.

From a materials science standpoint, the predicted microstructures retain the essential features relevant to expert analysis, including grain boundary topology, relative grain sizes, and physically consistent junction configurations, even at evaluation points where pixel-wise metrics indicate moderate degradation. This suggests that the predictions remain suitable for qualitative microstructural assessment and process optimization applications where exact boundary positions are less critical than overall structural characteristics.

Despite these results, there is further space for improvement. Future work may address the error accumulation during extended isothermal holds through modifications to the autoregressive prediction strategy, such as periodic incorporation of intermediate reference states or integration of physics-informed constraints. The demonstrated capability of thermal conditioning through FiLM to generalize across thermal regimes provides a foundation for extending the DL framework to the full range of industrial heat treatment processes.


\section*{Acknowledgments}
The authors thank ArcelorMittal, Aperam, Aubert \&
Duval, CEA, Constellium, Framatome, and Safran companies and the ANR for their financial support through the DIGIMU consortium and RealIMotion ANR Industrial Chair (Grant No. ANR-22-CHIN-0003).

\section*{Data Availability}
Data will be made available on request.

\section*{Conflicts of Interest}
The authors declare that they have no known competing financial interests or personal relationships that could have appeared to influence the work reported in this paper.

\section*{Declaration of generative AI and AI-assisted technologies in the manuscript preparation process}
During the preparation of this work the author(s) used ChatGPT in order to perform language editing and paraphrasing. After using this tool/service, the author(s) reviewed and edited the content as needed and take(s) full responsibility for the content of the published article.

\bibliographystyle{elsarticle-num}
\bibliography{references}


\clearpage

\setcounter{figure}{0}
\setcounter{table}{0}
\renewcommand{\thefigure}{S\arabic{figure}}
\renewcommand{\thetable}{S\arabic{table}}

\section*{Supplementary Material}


\begin{figure}[htbp]
    \centering
    \begin{subfigure}[t]{0.30\linewidth}
    \centering
    \includegraphics[width=\linewidth]{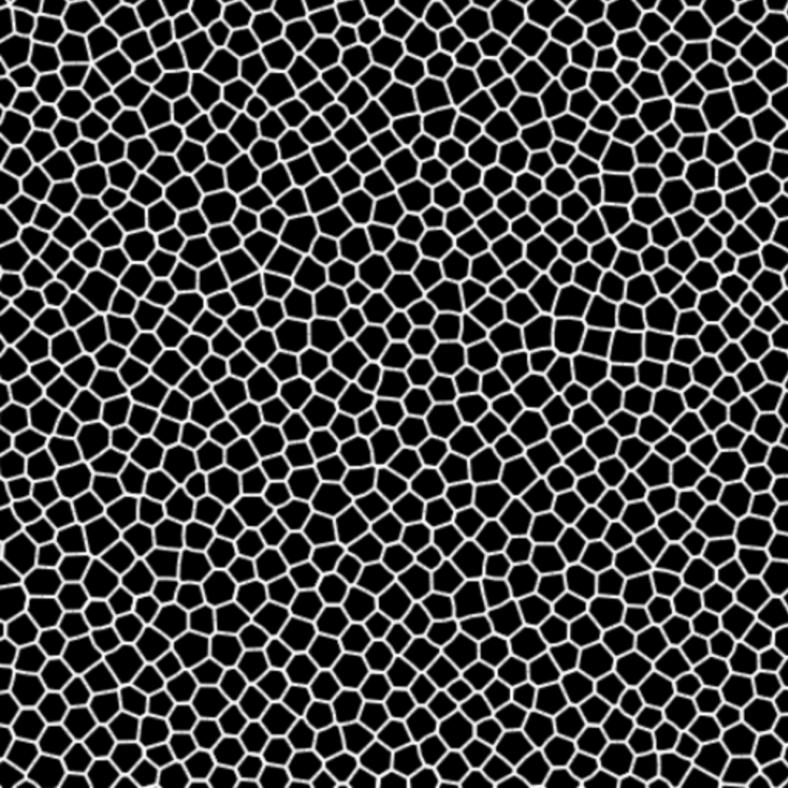}
    \caption{}
    \end{subfigure}
    \hfill
    \begin{subfigure}[t]{0.30\linewidth}
    \centering
    \includegraphics[width=\linewidth]{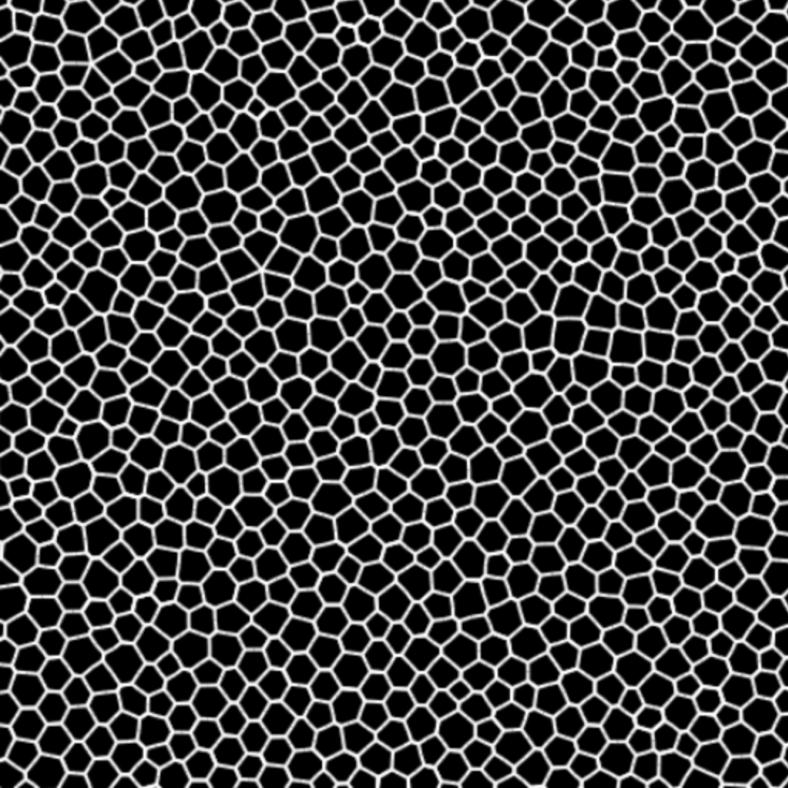}
    \caption{}
    \end{subfigure}
    \hfill
    \begin{subfigure}[t]{0.30\linewidth}
    \centering
    \includegraphics[width=\linewidth]{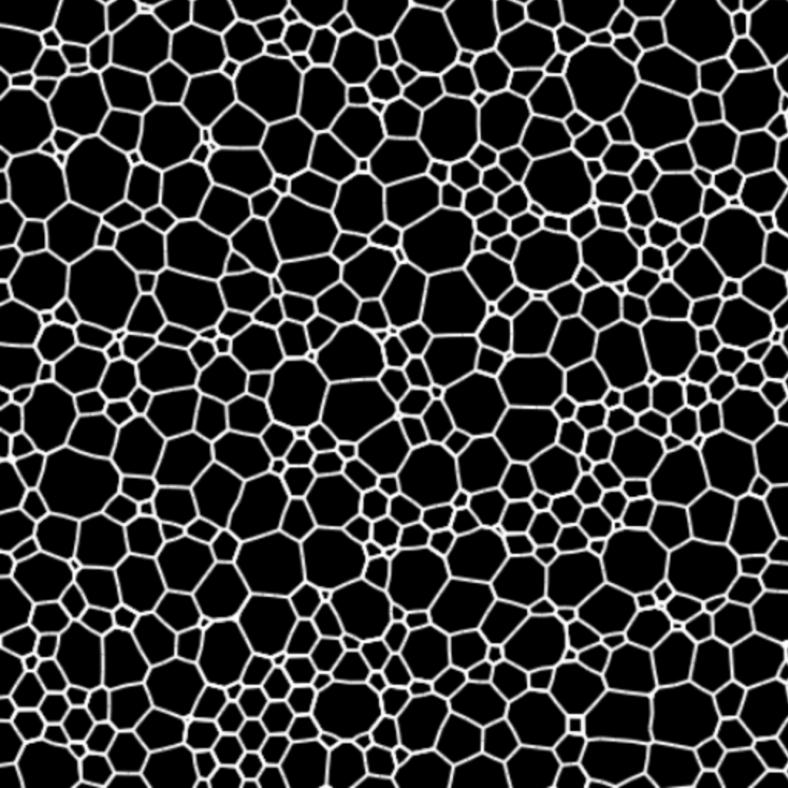}
    \caption{}
    \end{subfigure}
    \caption{Reference microstructure states for Scenario 1 simulated with the TRM: (a) initial state ($t = \SI{0}{\minute}$); (b) end of heating phase ($t = \SI{18}{\minute}$); (c) end of cooling phase ($t = \SI{59}{\minute}$).}
    \label{fig:s1_scenario1}
\end{figure}

\begin{figure}[htbp]
    \centering
    \begin{subfigure}[t]{0.30\linewidth}
    \centering
    \includegraphics[width=\linewidth]{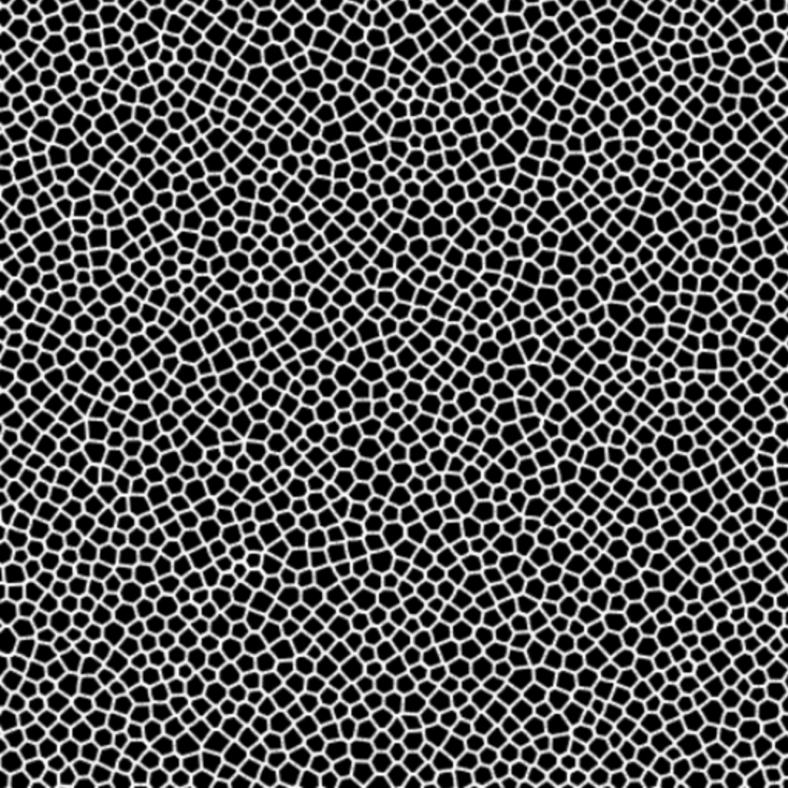}
    \caption{}
    \end{subfigure}
    \hfill
    \begin{subfigure}[t]{0.30\linewidth}
    \centering
    \includegraphics[width=\linewidth]{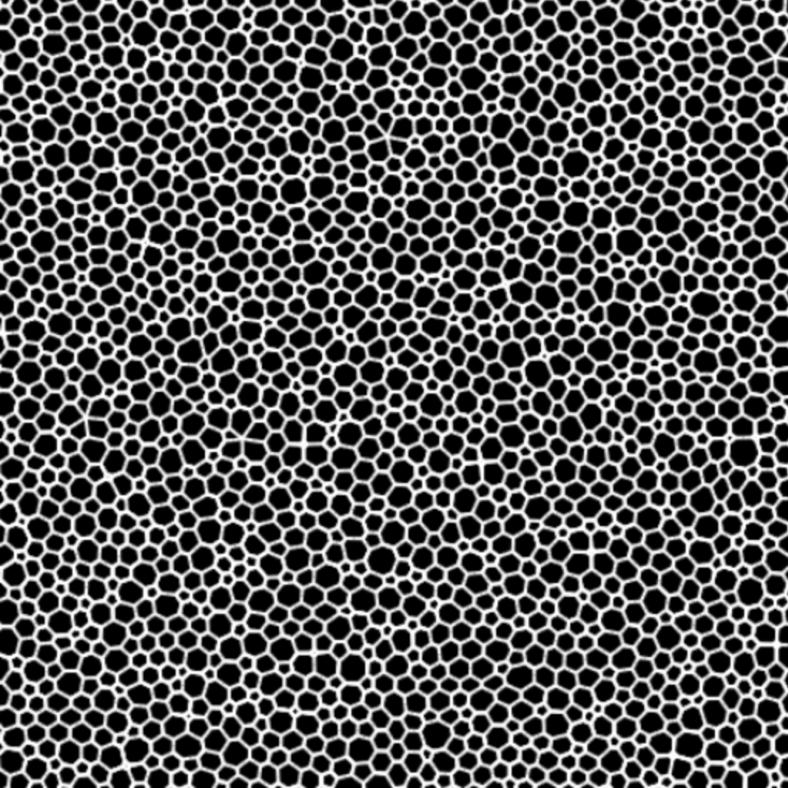}
    \caption{}
    \end{subfigure}
    \hfill
    \begin{subfigure}[t]{0.30\linewidth}
    \centering
    \includegraphics[width=\linewidth]{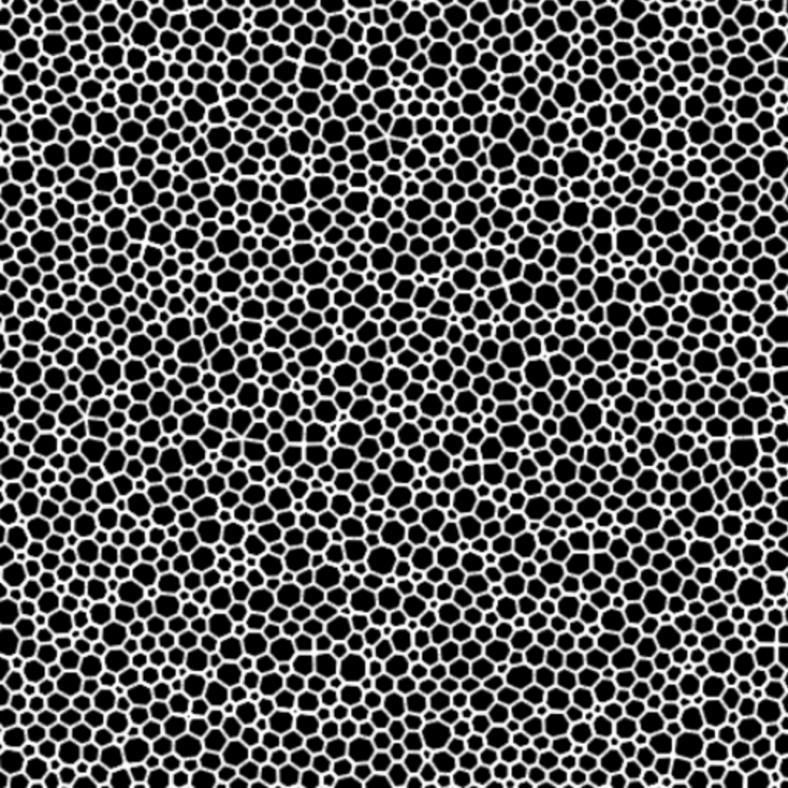}
    \caption{}
    \end{subfigure}
    \caption{Reference microstructure states for Scenario 2 simulated with the TRM: (a) initial state ($t = \SI{0}{\minute}$); (b) start of cooling phase ($t = \SI{26}{\minute}$); (c) end of cooling phase ($t = \SI{59}{\minute}$).}
    \label{fig:s1_scenario2}
\end{figure}

\begin{figure}[htbp]
    \centering
    \begin{subfigure}[t]{0.30\linewidth}
    \centering
    \includegraphics[width=\linewidth]{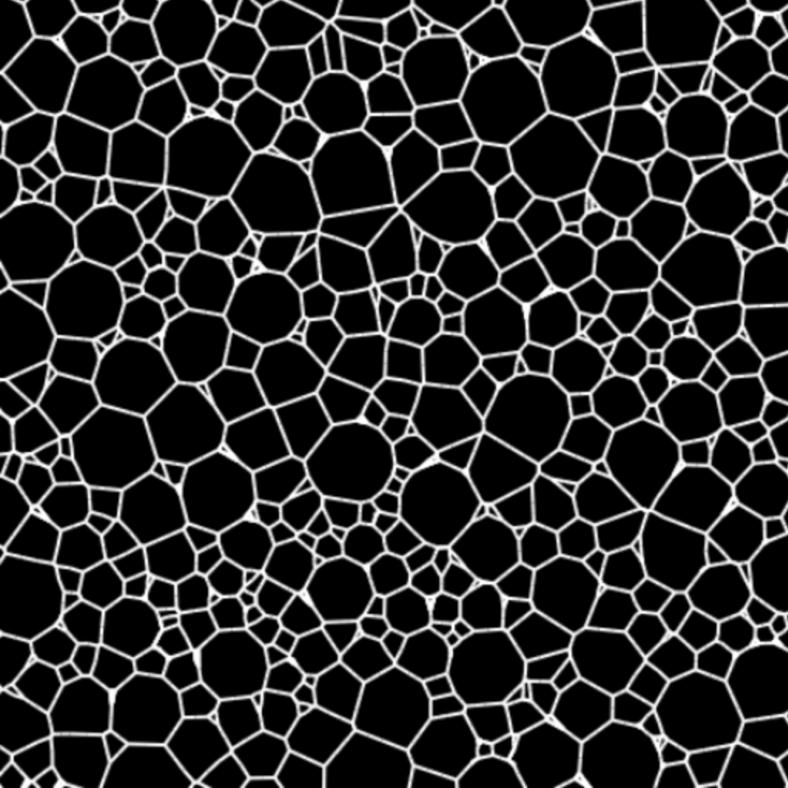}
    \caption{}
    \end{subfigure}
    \hfill
    \begin{subfigure}[t]{0.30\linewidth}
    \centering
    \includegraphics[width=\linewidth]{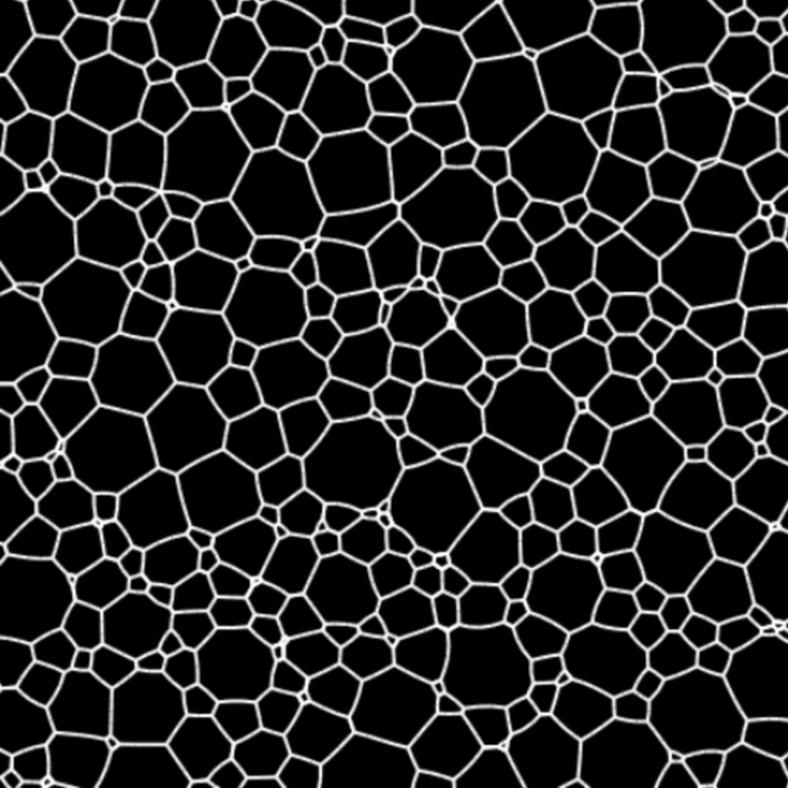}
    \caption{}
    \end{subfigure}
    \hfill
    \begin{subfigure}[t]{0.30\linewidth}
    \centering
    \includegraphics[width=\linewidth]{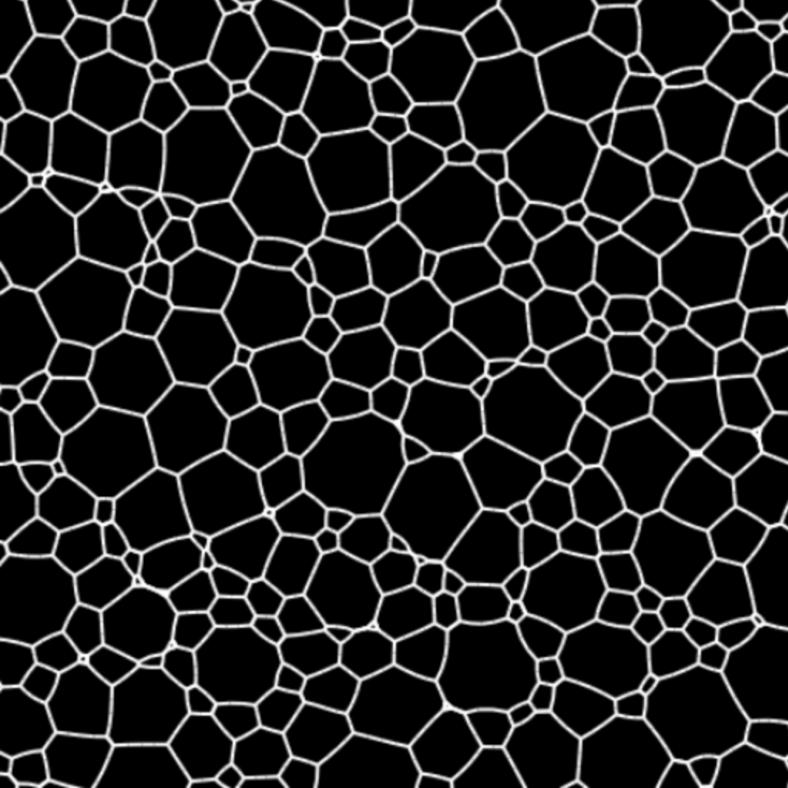}
    \caption{}
    \end{subfigure}

    \vspace{0.75em}

    \begin{subfigure}[t]{0.30\linewidth}
    \centering
    \includegraphics[width=\linewidth]{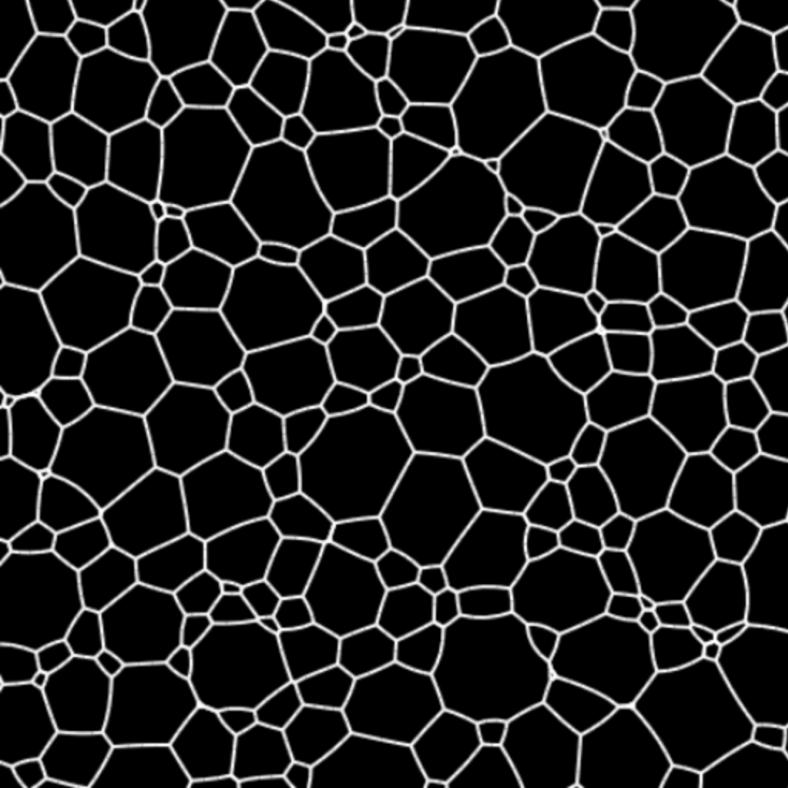}
    \caption{}
    \end{subfigure}
    \hspace{0.02\linewidth}
    \begin{subfigure}[t]{0.30\linewidth}
    \centering
    \includegraphics[width=\linewidth]{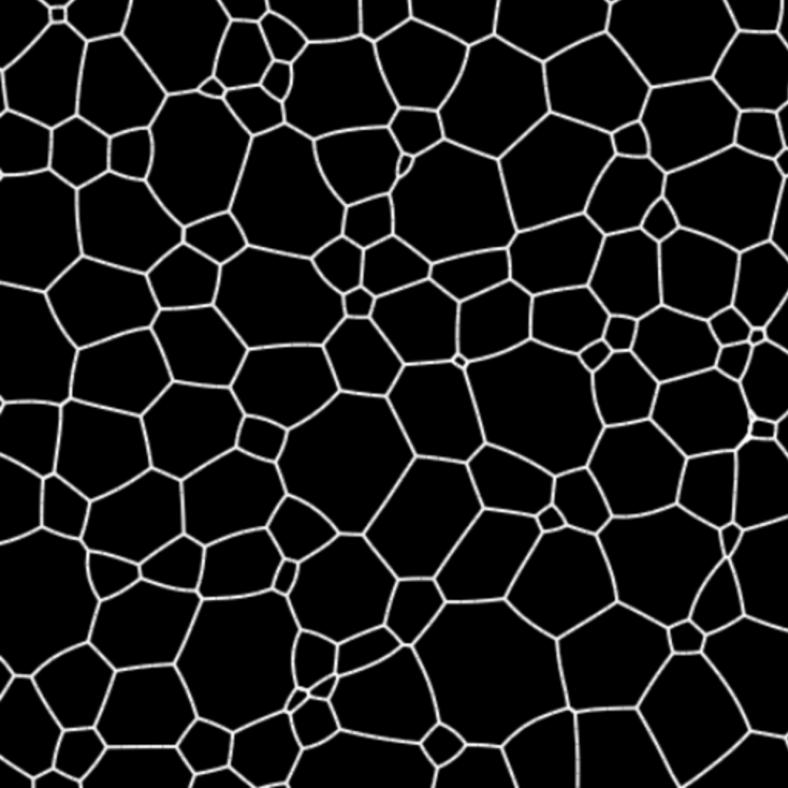}
    \caption{}
    \end{subfigure}
    \caption{Reference microstructure states for Scenario 3 simulated with the TRM: (a) initial state ($t = \SI{0}{\minute}$); (b) end of first heating ($t = \SI{17}{\minute}$); (c) end of partial cooling ($t = \SI{25}{\minute}$); (d) end of reheating ($t = \SI{32}{\minute}$); (e) end of complete thermal cycle ($t = \SI{59}{\minute}$).}
    \label{fig:s1_scenario3}
\end{figure}


\begin{figure}[htbp]
    \centering
    \begin{subfigure}[t]{0.48\linewidth}
    \centering
    \includegraphics[width=\linewidth]{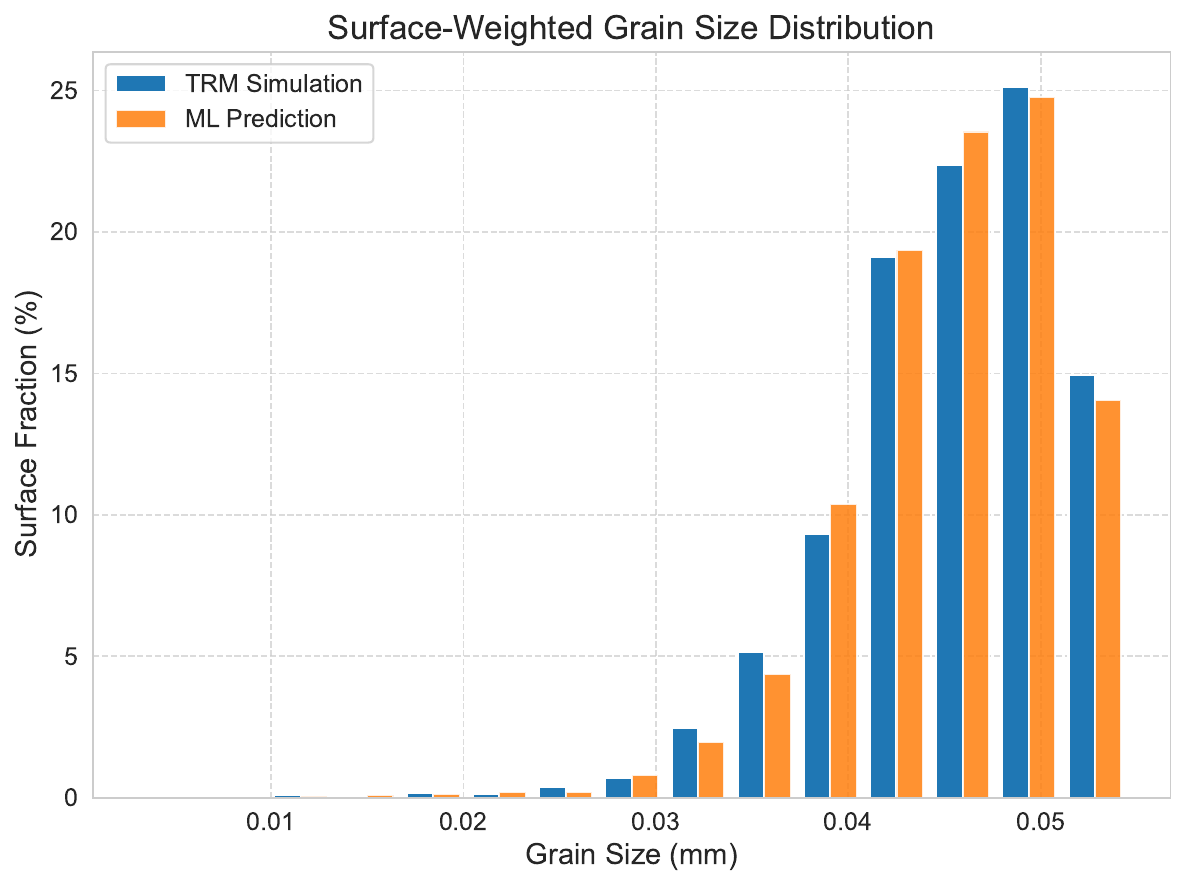}
    \caption{}
    \end{subfigure}
    \hfill
    \begin{subfigure}[t]{0.48\linewidth}
    \centering
    \includegraphics[width=\linewidth]{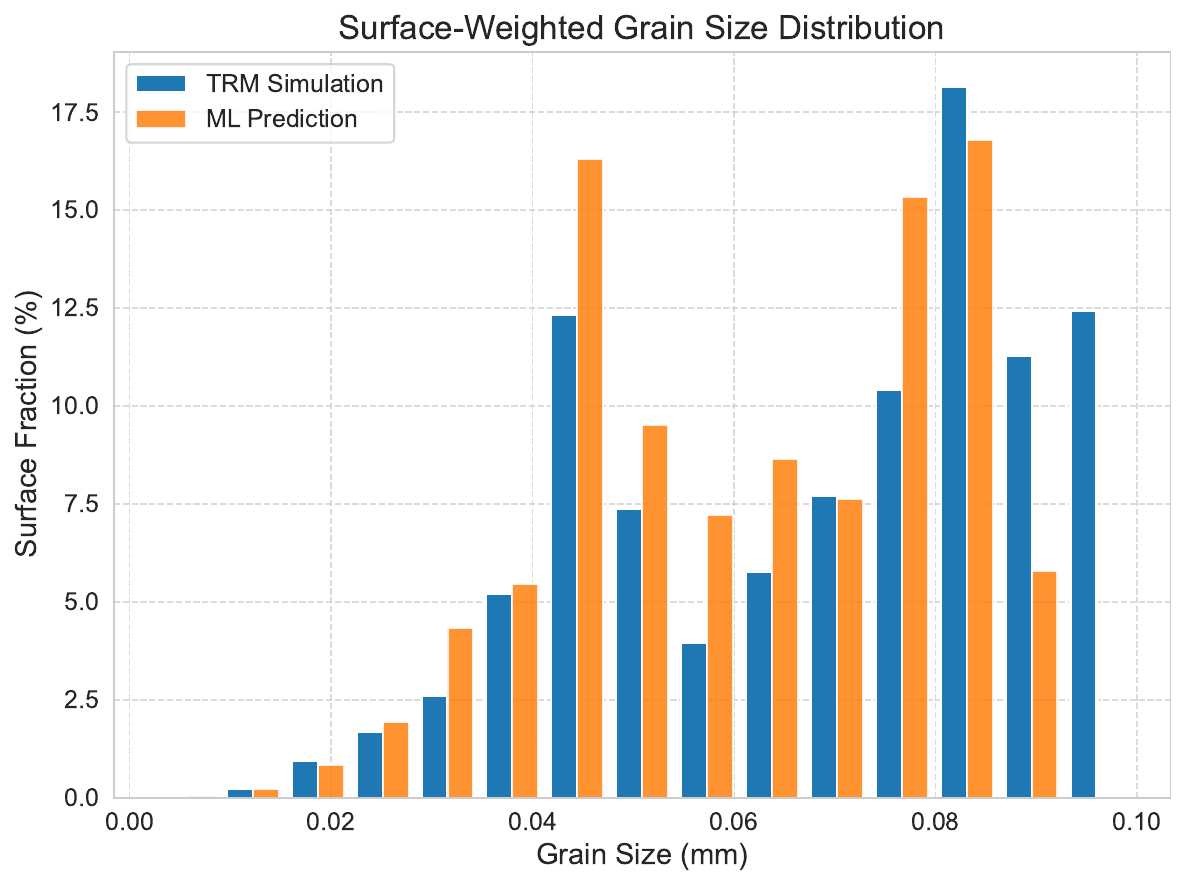}
    \caption{}
    \end{subfigure}

    \vspace{0.5em}

    \begin{subfigure}[t]{0.48\linewidth}
    \centering
    \includegraphics[width=\linewidth]{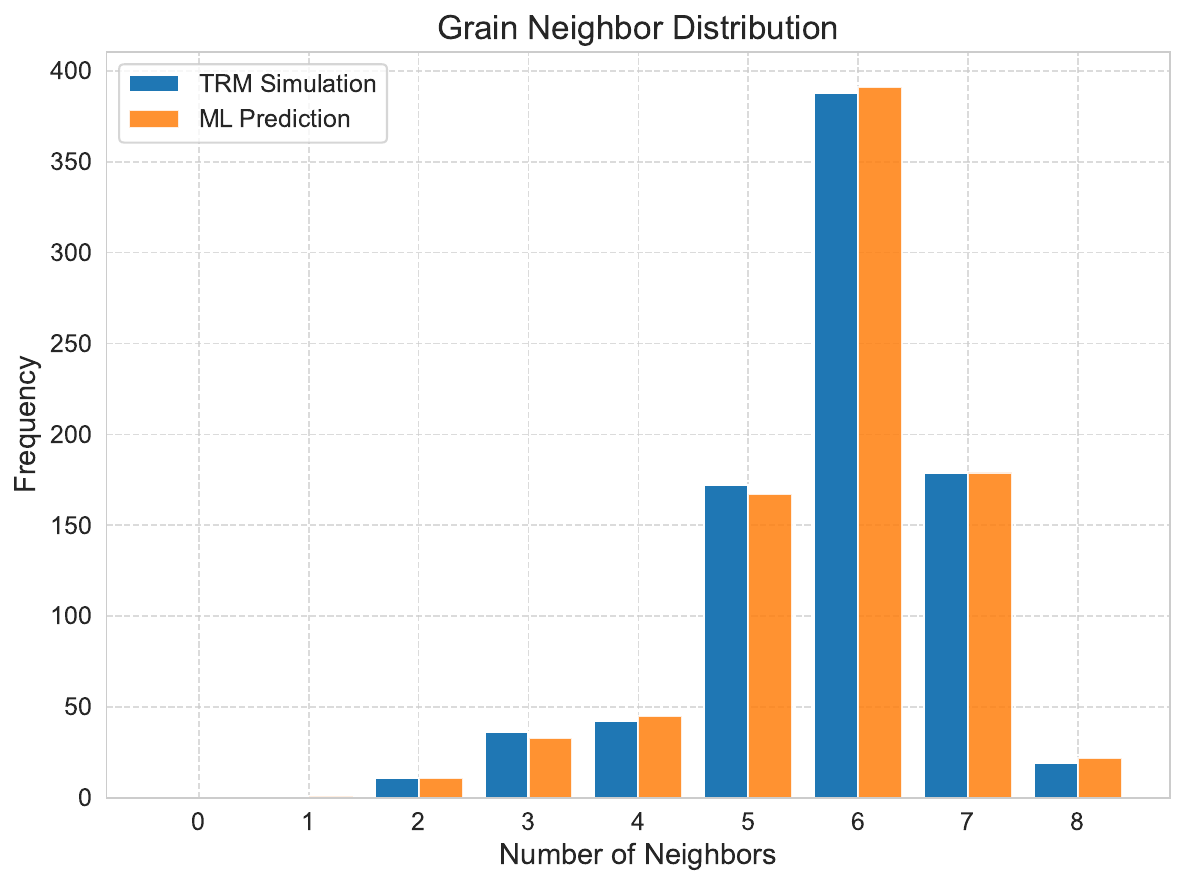}
    \caption{}
    \end{subfigure}
    \hfill
    \begin{subfigure}[t]{0.48\linewidth}
    \centering
    \includegraphics[width=\linewidth]{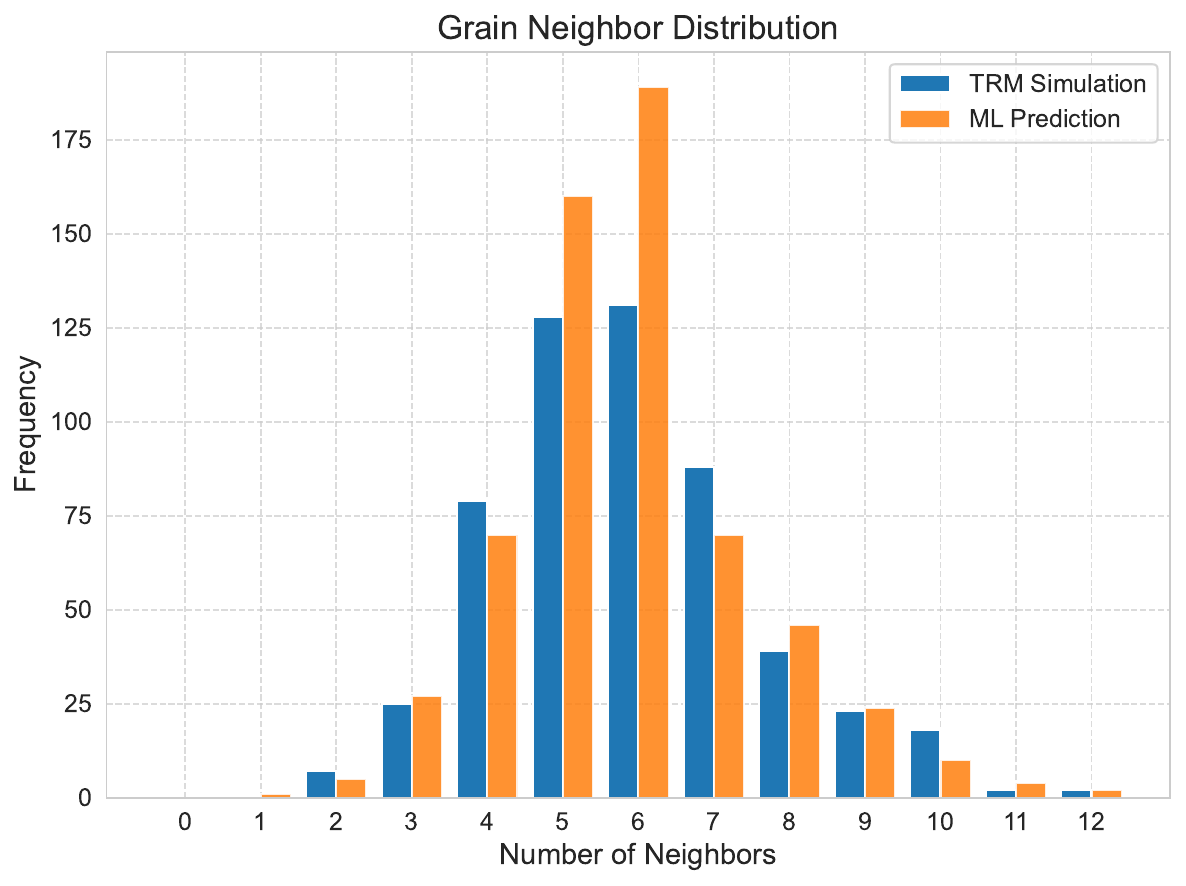}
    \caption{}
    \end{subfigure}
    \caption{Distributions comparison between predicted and ground truth for Scenario 1: (a, b) surface-weighted ECR distributions; (c, d) grain neighbor count distributions at $t = \SI{18}{\minute}$ and $t = \SI{59}{\minute}$, respectively.}
    \label{fig:s2_dist_scenario1}
\end{figure}


\begin{figure}[htbp]
    \centering
    \begin{subfigure}[t]{0.48\linewidth}
    \centering
    \includegraphics[width=\linewidth]{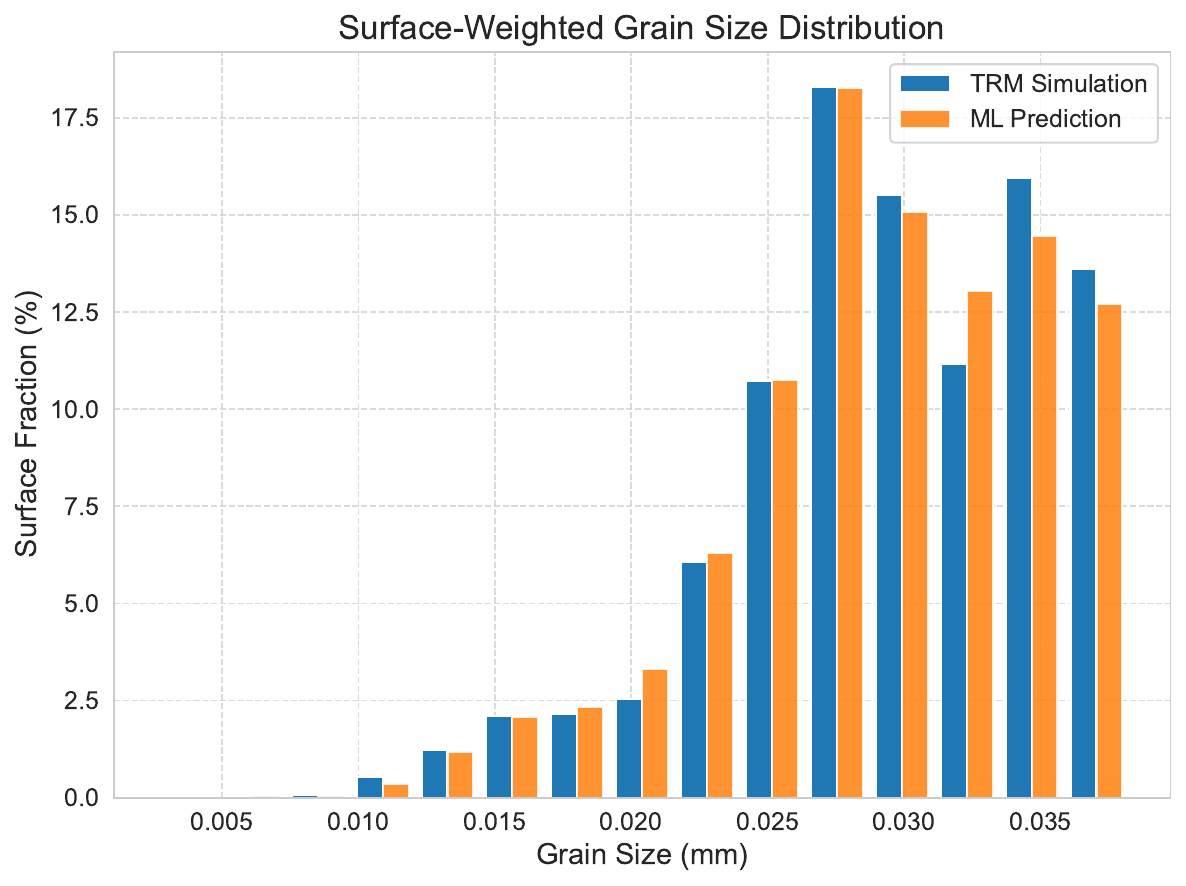}
    \caption{}
    \end{subfigure}
    \hfill
    \begin{subfigure}[t]{0.48\linewidth}
    \centering
    \includegraphics[width=\linewidth]{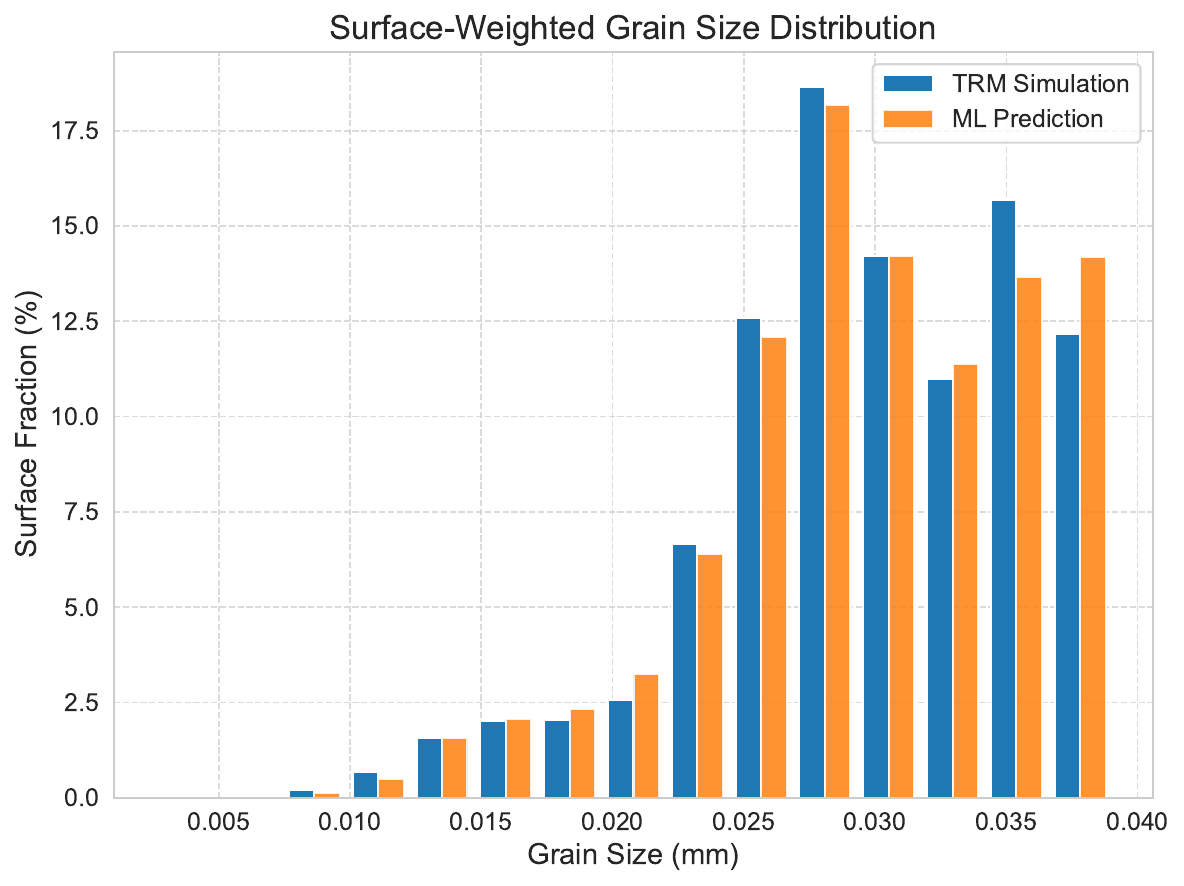}
    \caption{}
    \end{subfigure}

    \vspace{0.5em}

    \begin{subfigure}[t]{0.48\linewidth}
    \centering
    \includegraphics[width=\linewidth]{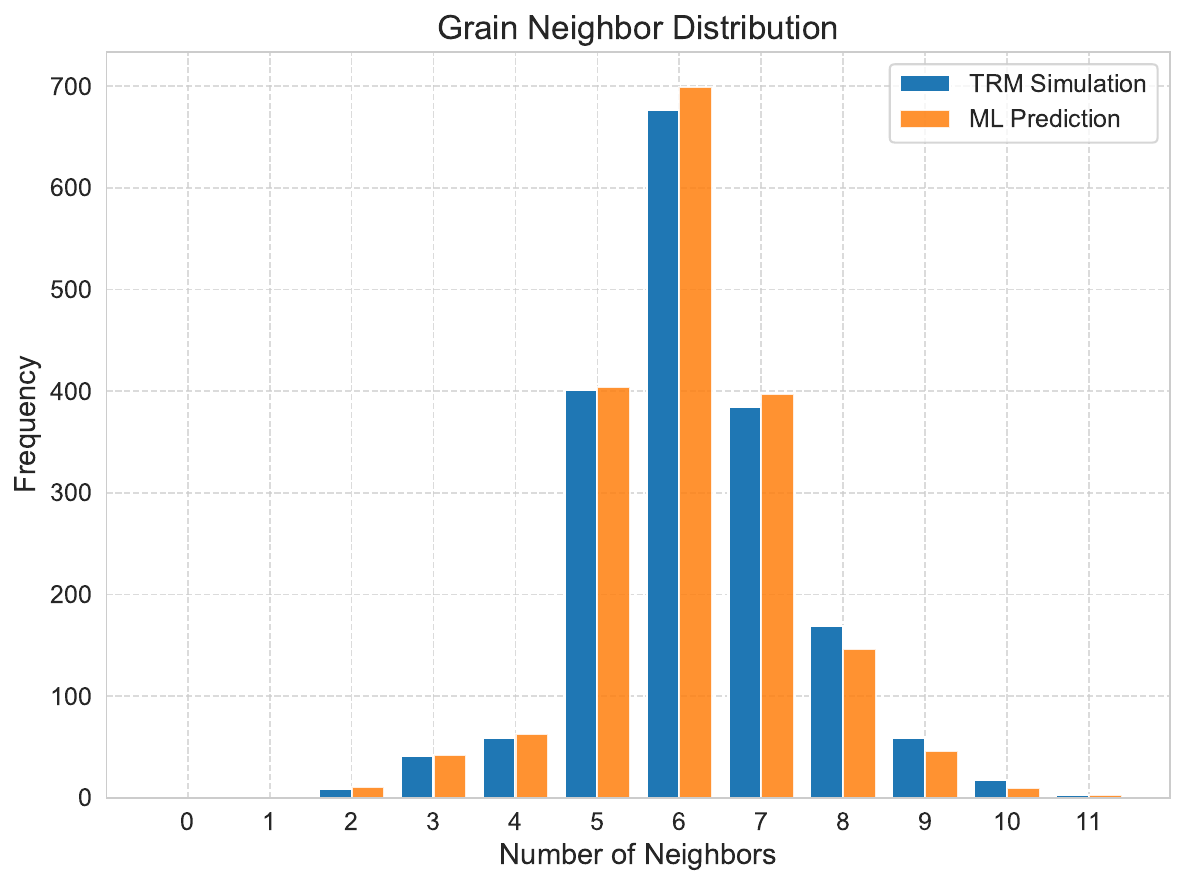}
    \caption{}
    \end{subfigure}
    \hfill
    \begin{subfigure}[t]{0.48\linewidth}
    \centering
    \includegraphics[width=\linewidth]{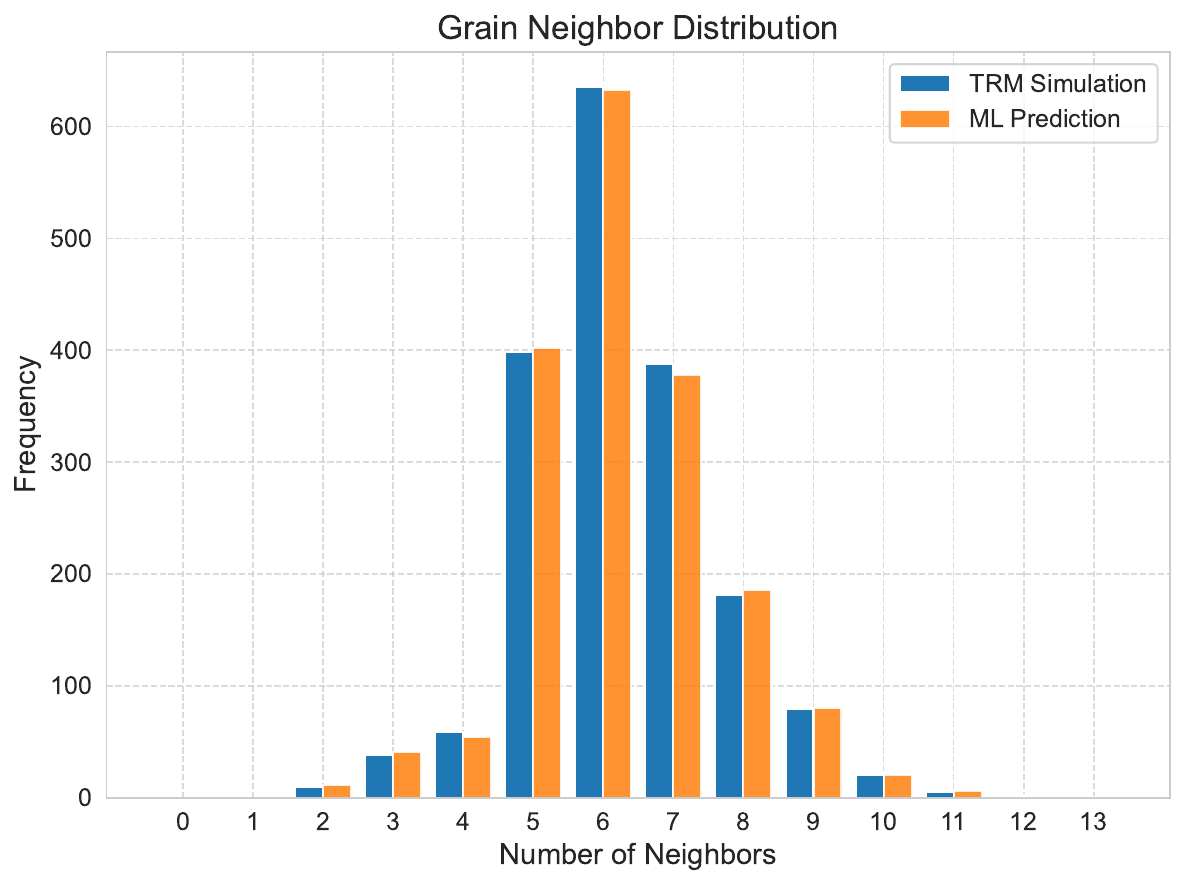}
    \caption{}
    \end{subfigure}
    \caption{Distributions comparison between predicted and ground truth for Scenario 2: (a, b) surface-weighted ECR distributions; (c, d) grain neighbor count distributions at $t = \SI{26}{\minute}$ and $t = \SI{59}{\minute}$, respectively.}
    \label{fig:s2_dist_scenario2}
\end{figure}


\begin{figure}[htbp]
    \centering
    \resizebox{0.85\textwidth}{!}{%

    \begin{minipage}{\textwidth}
    \centering
    \begin{subfigure}[t]{0.43\linewidth}
    \centering
    \includegraphics[width=\linewidth]{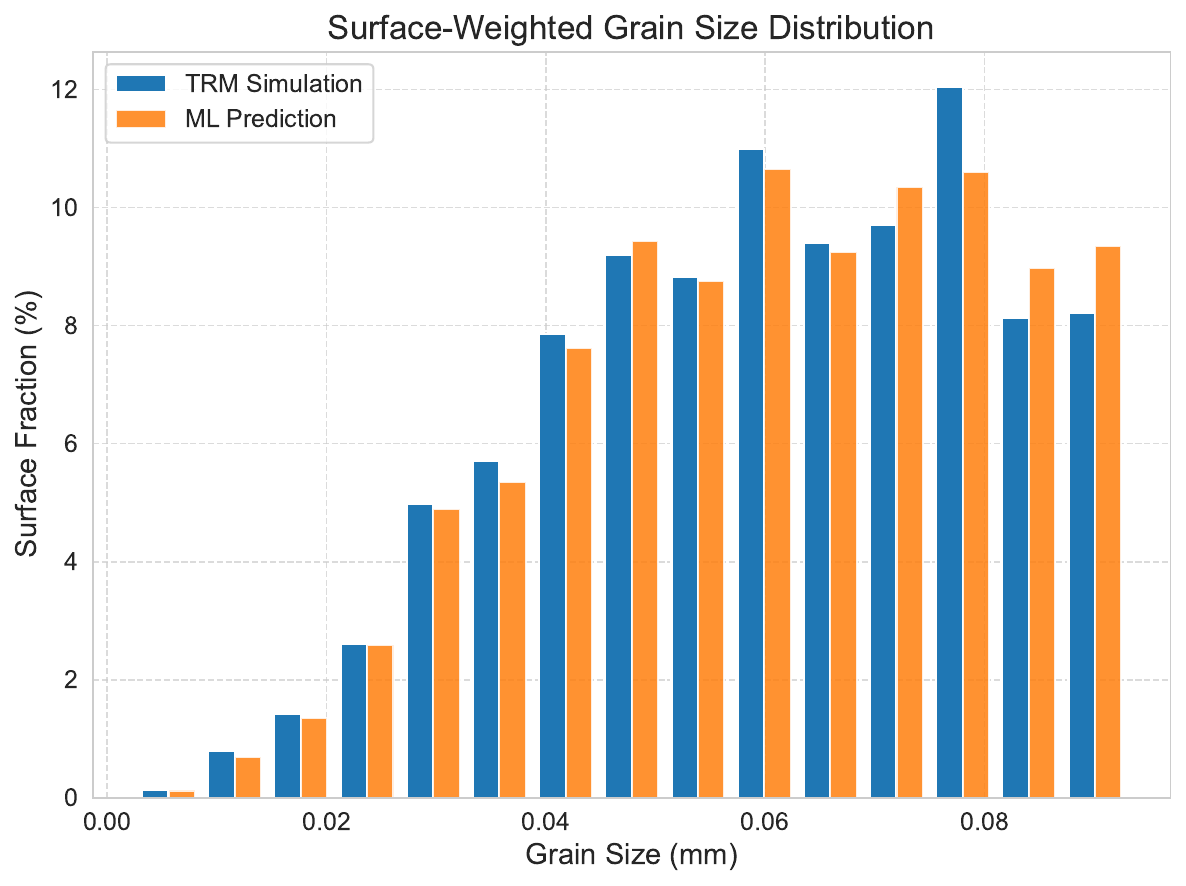}
    \caption{}
    \end{subfigure}
    \hfill
    \begin{subfigure}[t]{0.43\linewidth}
    \centering
    \includegraphics[width=\linewidth]{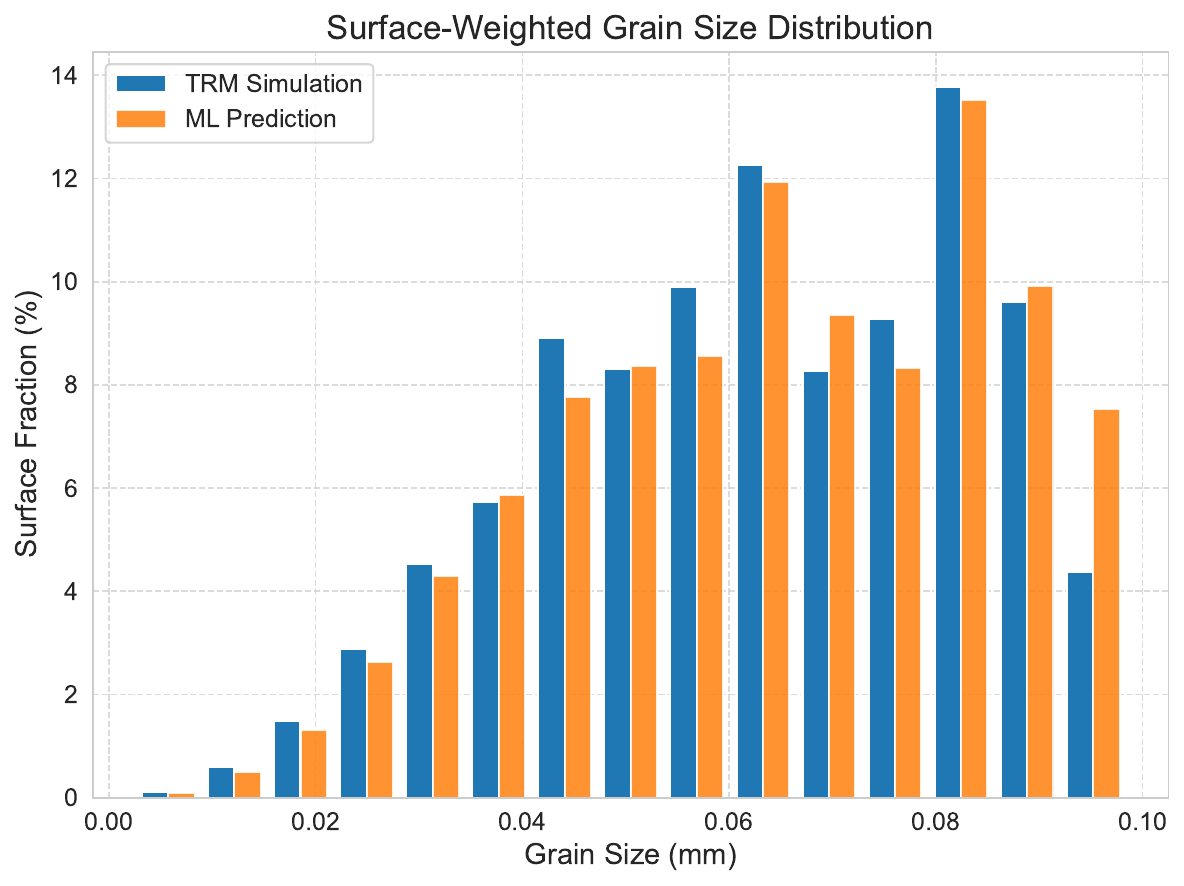}
    \caption{}
    \end{subfigure}

    \vspace{0.5em}

    \begin{subfigure}[t]{0.43\linewidth}
    \centering
    \includegraphics[width=\linewidth]{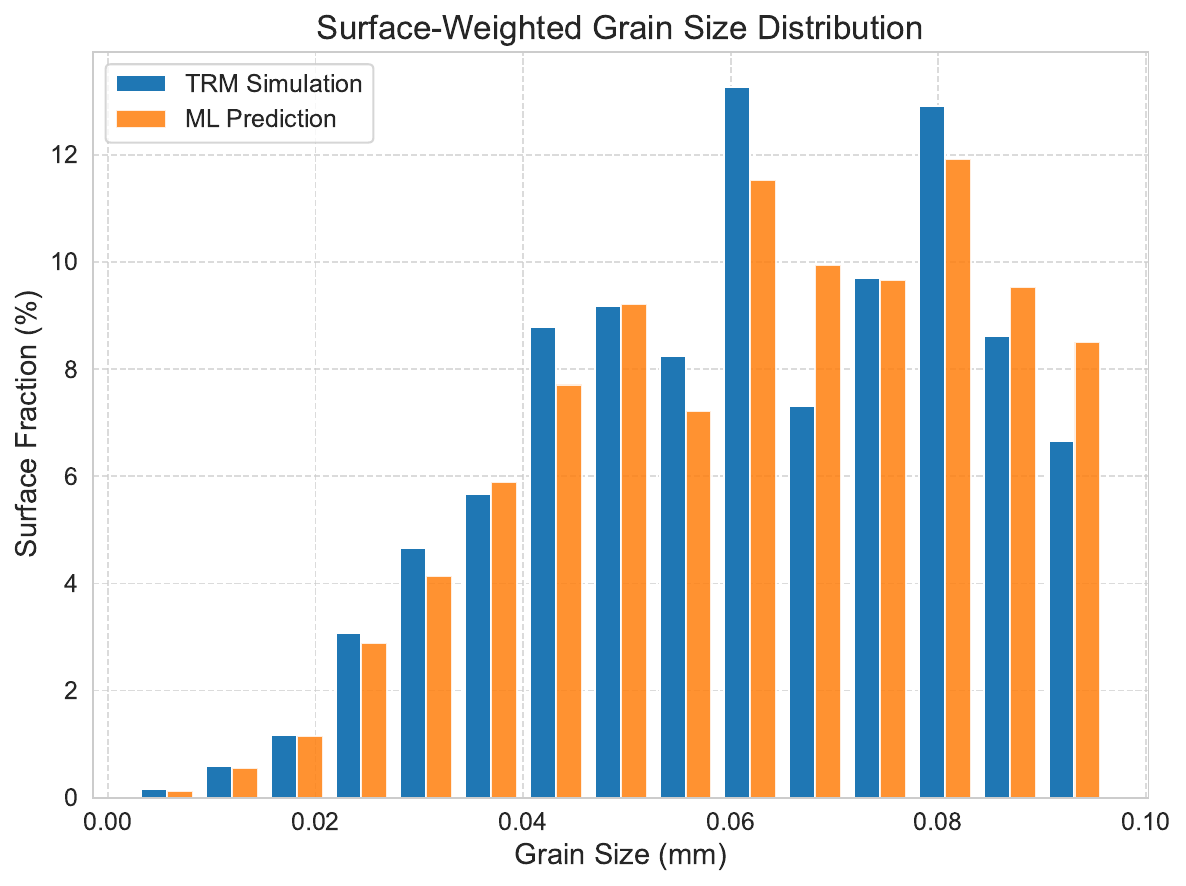}
    \caption{}
    \end{subfigure}
    \hfill
    \begin{subfigure}[t]{0.43\linewidth}
    \centering
    \includegraphics[width=\linewidth]{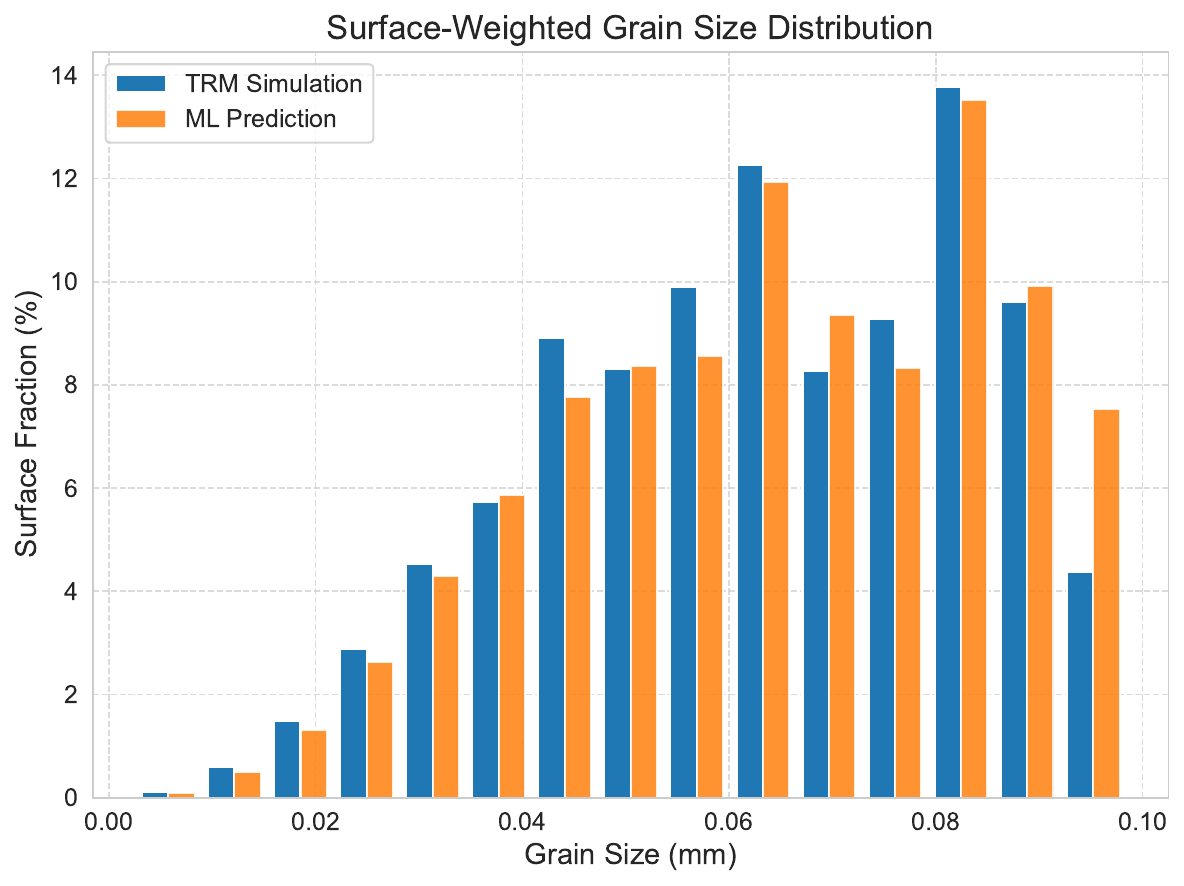}
    \caption{}
    \end{subfigure}

    \vspace{0.5em}

    \begin{subfigure}[t]{0.43\linewidth}
    \centering
    \includegraphics[width=\linewidth]{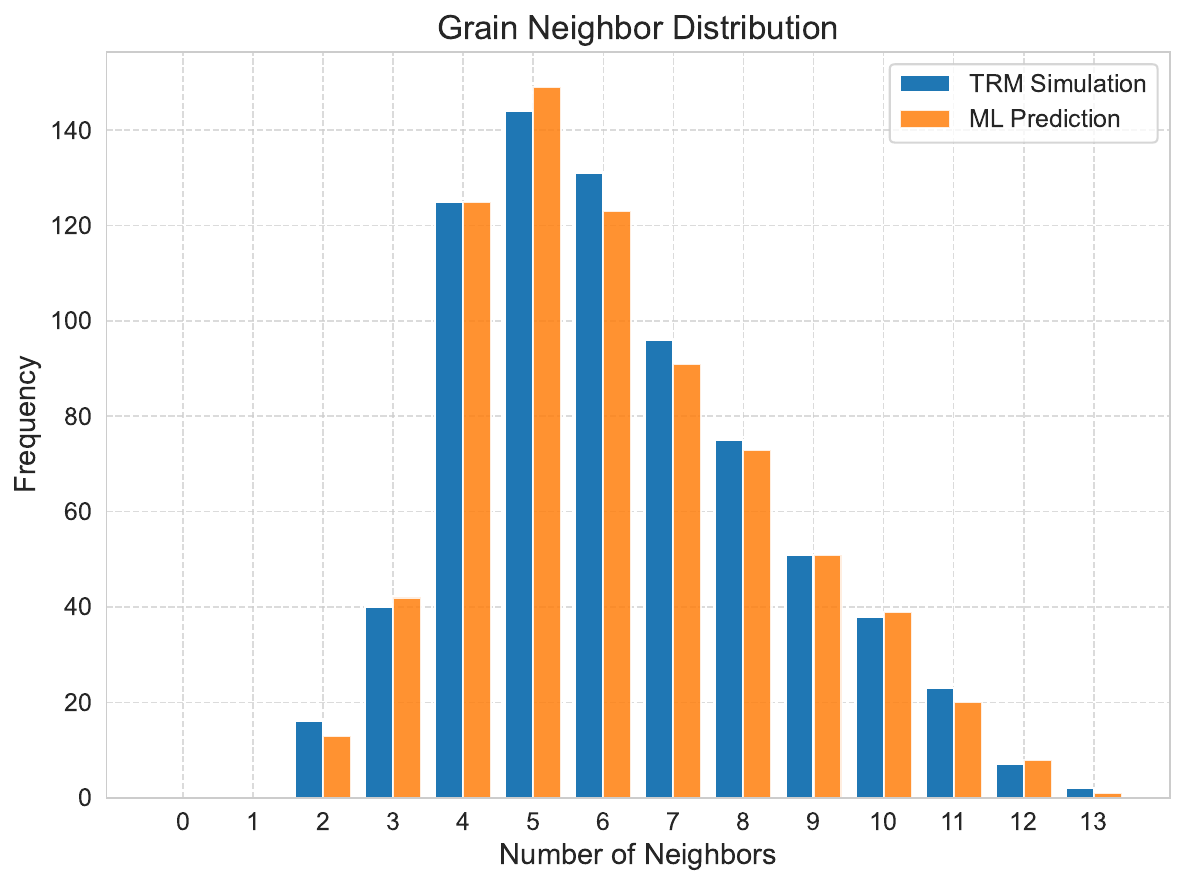}
    \caption{}
    \end{subfigure}
    \hfill
    \begin{subfigure}[t]{0.43\linewidth}
    \centering
    \includegraphics[width=\linewidth]{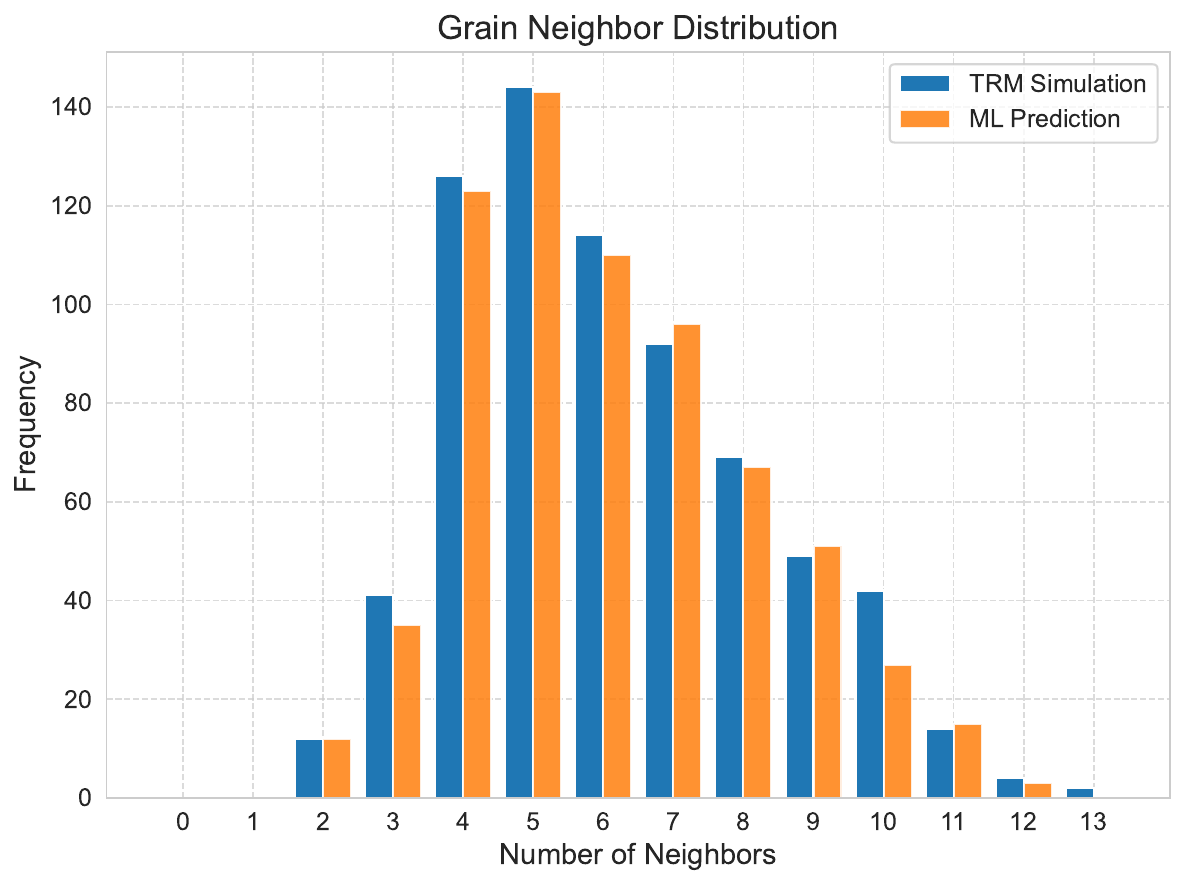}
    \caption{}
    \end{subfigure}

    \vspace{0.5em}

    \begin{subfigure}[t]{0.43\linewidth}
    \centering
    \includegraphics[width=\linewidth]{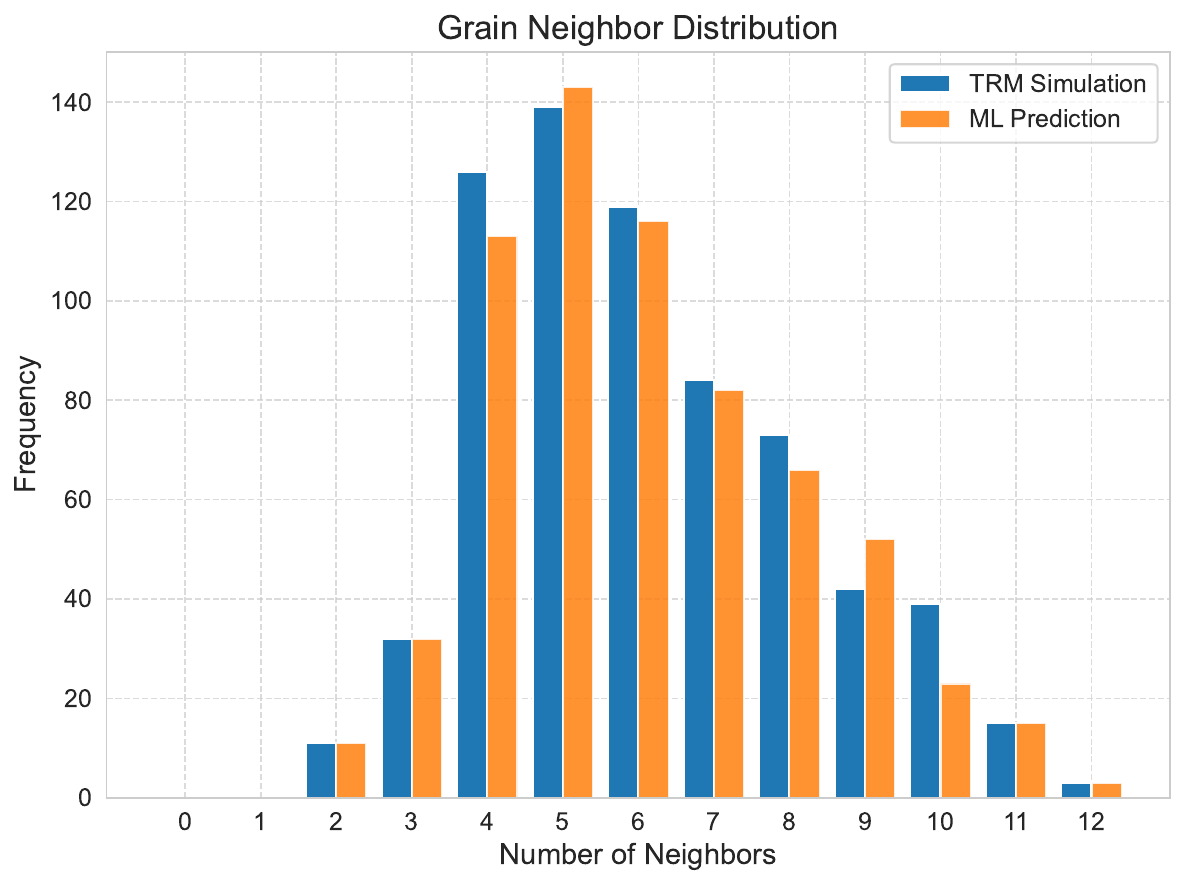}
    \caption{}
    \end{subfigure}
    \hfill
    \begin{subfigure}[t]{0.43\linewidth}
    \centering
    \includegraphics[width=\linewidth]{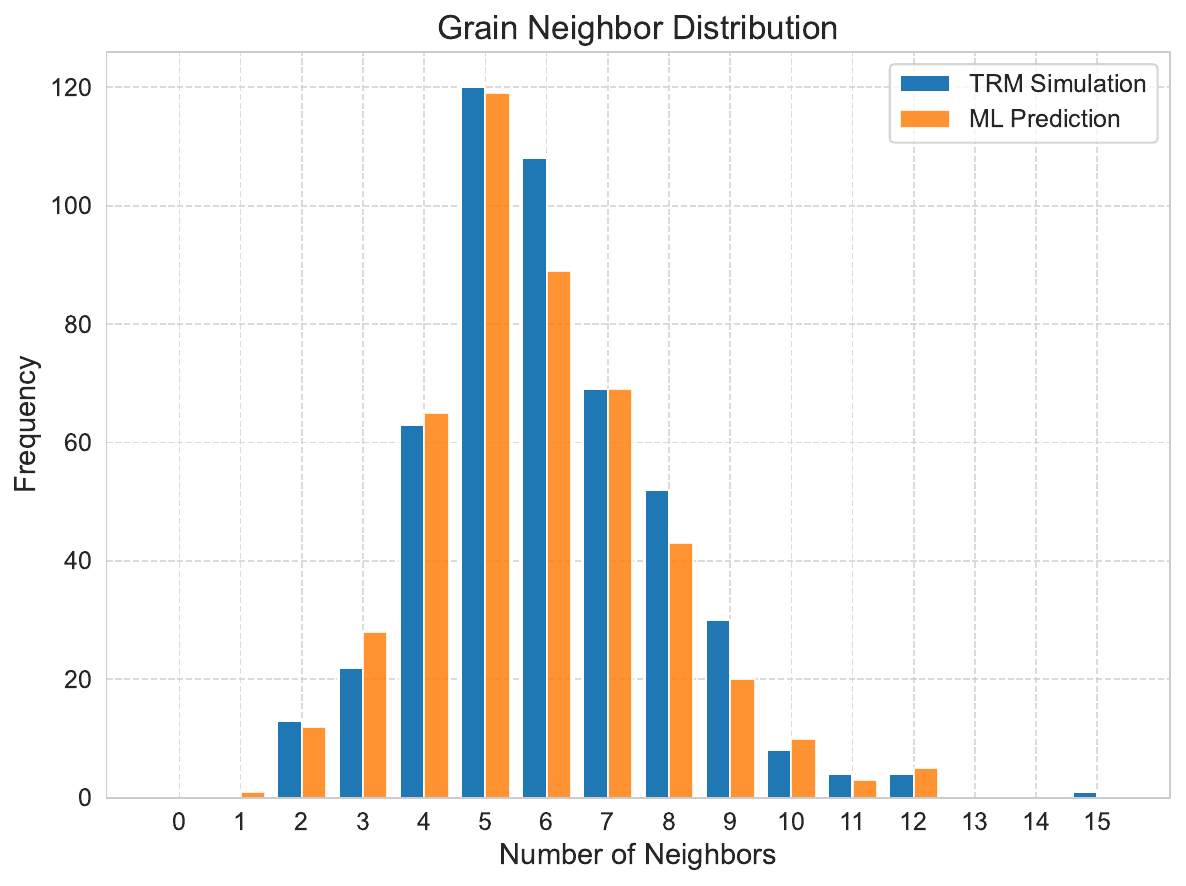}
    \caption{}
    \end{subfigure}
    \end{minipage}%

    }
    \caption{Distributions comparison between predicted and ground truth for Scenario 3: (a--d) surface-weighted ECR distributions; (e--h) grain neighbor count distributions at $t = \SI{17}{\minute}$, $\SI{25}{\minute}$, $\SI{32}{\minute}$, and $\SI{59}{\minute}$, respectively.}
    \label{fig:s2_dist_scenario3}
\end{figure}

\end{document}